\documentclass[a4paper,11pt]{article}
\usepackage{graphicx}
\usepackage{amsmath, amsthm, amsfonts, amssymb, bbm}
\usepackage[mathscr]{eucal}

\theoremstyle{plain}
\newtheorem{lemma}{Lemma}[section]

\theoremstyle{remark}
\newtheorem{remark}{Remark}[section]

\newcommand{\rhs}{r.h.s.\ }

\newcommand{\ud}{\mathrm{d}}
\newcommand{\del}{\partial}
\newcommand{\bra}[1]{\langle #1 |}
\newcommand{\ket}[1]{| #1 \rangle}

\newcommand{\tr}{\mathrm{tr}}

\DeclareMathOperator{\si}{si}
\DeclareMathOperator{\ci}{ci}

\newcommand{\V}[1]{\mathbf{#1}}

\newcommand{\order}{\mathcal{O}}

\newcommand{\R}{\mathbb{R}}
\newcommand{\1}{\mathbbm{1}}

\newcommand{\betrag}[1]{\left| #1 \right|}

\newcommand{\pv}[1]{\mathrm{P} \frac{1}{#1}}

\renewcommand{\slash}[2][4]{\ensuremath{\rlap{\raisebox{1pt}{$\mspace{#1mu}/$}}#2}}

\newcommand{\Dslash}{\slash{D}}

\newcommand{\kslash}{\slash[2]{k}}

\newcommand{\dotalpha}{{\dot \alpha}}
\newcommand{\dotbeta}{{\dot \beta}}

\numberwithin{equation}{section}

\begin{document}

\title{Noncommutative (supersymmetric) electrodynamics in the Yang-Feldman formalism}
\date{\today}
\author{ 
Jochen Zahn \\
Courant Research Centre ``Higher Order Structures'' \\
University of G\"ottingen \\
Bunsenstra{\ss}e 3--5, D-37073 G\"ottingen, Germany \\
jzahn@uni-math.gwdg.de
}

\maketitle

\begin{abstract}
We study quantum electrodynamics on the noncommutative Minkowski space in the Yang-Feldman formalism. Local observables are defined by using covariant coordinates. We compute the two-point function of the interacting field strength to second order and find the infrared divergent terms already known from computations using the so-called modified Feynman rules. It is shown that these lead to nonlocal renormalization ambiguities. Also new nonlocal divergences stemming from the covariant coordinates are found. Furthermore, we study the supersymmetric extension of the model. For this, the supersymmetric generalization of the covariant coordinates is introduced. We find that the nonlocal divergences cancel. At the one-loop level, the only effect of noncommutativity is then a momentum-depenent field strength normalization. We interpret it as an acausal effect and show that its range is independent of the noncommutativity scale.
\end{abstract}

\section{Introduction}

Using semiclassical arguments, Doplicher, Fredenhagen and Roberts \cite{dfr} showed that the interplay of general relativity and quantum theory leads to restrictions on the localizability of events in different space-time coordinates (space-time uncertainty relations). They proposed to implement these by using coordinates $q^\mu$ fulfilling commutation relations
\begin{equation}
\label{eq:q_comm}
 [q^\mu, q^\nu] = i \Theta^{\mu \nu}.
\end{equation}
Here $\Theta$ is an antisymmetric matrix and an element of $\Sigma$, which is the orbit of
\begin{equation}
\label{eq:sigma_0}
 \Theta_0 = \lambda_{\text{nc}}^2 \begin{pmatrix} 0 & - \mathbbm{1}_2 \\ \mathbbm{1}_2 & 0   \end{pmatrix}
\end{equation}
under the Lorentz group\footnote{Strictly speaking, one should replace $\Theta^{\mu \nu}$ in (\ref{eq:q_comm}) by a central operator $Q^{\mu \nu}$ whose joint spectrum is $\Sigma$. But for the present purposes, it suffices to consider a fixed $\Theta \in \Sigma$, as long as one keeps in mind that it transforms as a tensor under Lorentz transformations.}. Here $\lambda_{\text{nc}}$ is a length scale that is typically identified with the Planck length. The elements of the noncommutative algebra $\mathcal{E}_\Theta$ generated by the $q^\mu$'s can then be thought of as functions on the noncommutative Minkowski space. Similar commutation relations appeared in the theory of open strings ending on D-branes in the presence of a background magnetic field~\cite{Schomerus, SW}. In this context, however, one can only arrive at space/space noncommutativity, i.e., $\Theta^{0 i} = 0$ \cite{EField}.

In recent years, a lot of effort has been invested in the construction and study of quantum field theories on the noncommutative Minkowski space.
There are three\footnote{Here, we do not consider the theories obtained by adding a Grosse-Wulkenhaar potential \cite{LangmannSzabo}. As shown in \cite{GW2d}, their Minkowski space version is badly divergent. We also do not consider the so-called $1/p^2$ theories \cite{p2}, and the models inspired by the twist approach of Wess and coworkers \cite{twist}.} main quantization methods used in the literature.

The \emph{modified Feynman rules} first proposed by Filk~\cite{Filk} can be obtained in a Euclidean path integral formalism. They amount to attaching a phase factor depending on the momenta to each vertex. At the one-loop level, there are then two possibilities: Either the phases cancel and the graph is exactly as in the commutative. This is the simplest example of a planar graph\footnote{In general, planarity is defined by the genus of the Riemann surface onto which the graph may be drawn \cite{chepelev}.}. Or they add up, in which case one obtains a nonplanar graph. In this graph, the incoming momentum $k$ (more precisely $((k \Theta)^{2})^{-1/2}$) serves as a UV regulator. Thus, these graphs are finite for nonvanishing $k$. However, problems appear if such a graph is embedded into a larger graph where one has to integrate over $k=0$. This is called the UV/IR-mixing problem, which may spoil the renormalizability~\cite{Minwalla}. Furthermore, the nonplanar graphs lead to a distortion of the dispersion relation~\cite{Matusis}. While these phenomena also appear in other approaches to NCQFT, there is one particular drawback of this approach: In the case of space-time noncommutativity $\Theta^{0i} \neq 0$, the modified Feynman rules are only valid in a Euclidean setting. The connection to Lorentzian signature is not clear \cite{DoroOWR}. 

The \emph{Hamiltonian approach} \cite{dfr, LiaoSibold} is applicable in a Lorentzian setting. Several propositions for a suitable Hamiltonian exist in the literature for the case of a scalar field theory. Some even lead to UV-finite models~\cite{UVfinite}. However, in the case of quantum electrodynamics, the Ward identities are not fulfilled already at tree level~\cite{Ohl}.

The \emph{Yang-Feldman approach} \cite{YF, BDFP02} is also adapted to the Lorentzian case. Since it directly uses the classical equations of motion, the classical symmetries are better preserved than in the Hamiltonian approach. Also in this approach the nonplanar graphs lead to a distortion of the dispersion relations~\cite{Quasiplanar, NCDispRel}.
A complete treatment of the UV/IR problem (including higher loop orders), analogous to the one presented in \cite{chepelev} for the Euclidean case, is still an open problem in this setting\footnote{
The difficulty is that in the Yang-Feldman approach one deals with two different propagators, which always have to be combined appropriately.}. In the present work, we are only considering the first nontrivial order, i.e., the one-loop level. It turns out that already at this order NCQED can not be renormalized by local counterterms.

One of the main effects of noncommutativity thus seems to be the distortion of the dispersion relations. The best experimental limits for such an effect probably exist for the photon. It is thus natural to consider a noncommutative version of QED. There, the field strength is given by
\begin{equation}
\label{eq:F}
 F_{\mu \nu} = \del_\mu A_\nu - \del_\nu A_\mu - i e [A_\mu, A_\nu].
\end{equation}
This looks like the field strength of a nonabelian gauge theory, but the commutator on the right hand side is now the commutator in the noncommutative algebra $\mathcal{E}_\Theta$. The action is then given by 
\begin{equation}
\label{eq:NCEDAction}
 S = - \frac{1}{4} \int \ud^4q \ F_{\mu \nu} F^{\mu \nu},
\end{equation}
where the integral is cyclic and the analog of the integral on the ordinary Minkowski space (for details on the noncommutative calculus, see below). Obviously, due to the nonlinear term in~(\ref{eq:F}), already pure electrodynamics on the noncommutative Minkowski space is interacting. 

There are two main approaches to NCQED\footnote{For a recent review on noncommutative gauge theories which also covers approaches that are inspired by the Grosse-Wulkenhaar model, we refer to~\cite{Blaschke}.}: In the approach via the Seiberg-Witten map~\cite{SW}, the gauge fields $A_\mu$ on the noncommutative Minkowski space are expressed as functions of gauge fields $\tilde A_\mu$ on the ordinary Minkowski space in a formal expansion in the noncommutativity parameter $\Theta$. One can then expand the action~(\ref{eq:NCEDAction}) in $\Theta$. At zeroth order in $\Theta$, one recovers the action of ordinary electrodynamics, while at first order in $\Theta$ one obtains terms cubic in the field strength. However, the validity of the expansion in $\Theta$ is not clear. Furthermore, it has been shown~\cite{Wulkenhaar} that NCQED in this setting is not renormalizable\footnote{However, this is only the case if fermions are included.}, already at first order in $\Theta$.

In the unexpanded approach, which was first studied in~\cite{Martin}, one treats NCQED in the same way as other noncommutative field theories, i.e., without expansion in $\Theta$. In the literature, it has mainly been treated with the modified Feynman rules. In~\cite{Ohl} the Hamiltonian approach was used and it was found that the Ward identity is violated already at tree level. In this paper, we are going to use the Yang-Feldman approach.

In~\cite{hayakawa} it was shown that in the unexpanded approach, in the setting of the modified Feynman rules, the non-planar part of the photon self-energy is of the form
\begin{equation}
\label{eq:SelfEnergy}
(2\pi)^2 \Pi^{\mu \nu}_{np}(k) = (g^{\mu \nu} k^2 - k^{\mu} k^{\nu}) \Sigma_1(k^2, (k \Theta)^2) + \tfrac{(k \Theta)^{\mu} (k \Theta)^{\nu}}{(k \Theta)^4} \Sigma_2(k^2, (k \Theta)^2)
\end{equation}
with
\begin{subequations}
\begin{align}
\label{eq:Hayakawa_Sigma_1}
 \Sigma_1(k^2, (k \Theta)^2) & = \tfrac{5}{3} \ln (\sqrt{k^2} \sqrt{- (k \Theta)^2} ) + \order(1) \\
\label{eq:Hayakawa_Sigma_2}
 \Sigma_2(k^2, (k \Theta)^2) & = 8 + \order(k^2 (k \Theta)^2).
\end{align}
\end{subequations}
The same result has been found in~\cite{KhozeTravaglini} with the background field method. From~(\ref{eq:Hayakawa_Sigma_2}) it is obvious that the second term in~(\ref{eq:SelfEnergy}) is quadratically IR-divergent. This was not expected, since the commutative theory is only logarithmically UV-divergent. In~\cite{Matusis} this was explained as follows: The underlying UV-divergence is quadratic by power-counting, only by invoking the Ward identity does it become logarithmic. In the nonplanar diagrams, however, the phase factor with the incoming momentum serves as a UV-regulator. Hence the quadratic IR-divergence in the incoming momentum. It was shown in~\cite{Ruiz} that this term is independent of the chosen gauge.

The strange second term in~(\ref{eq:SelfEnergy}) is usually interpreted as a severe distortion of the dispersion relation~\cite{Matusis}, leading to tachyonic modes\footnote{A different interpretation was put forward in \cite{UVIRemergent} in the context of the emergent gravity scenario, where it is argued that these modes are gravitational degrees of freedom.} \cite{Ruiz}. In~\cite{GraciaBondia} it was argued that in the case of space/space noncommutativity, it leads to ill-defined terms in the effective action. Here we adopt the point of view put forward in~\cite{DoroOWR}: In the full expression for the two-point function, $\frac{(k\Theta)^\mu (k\Theta)^\mu}{(k \Theta)^4}$ is multiplied with $\theta(k_0) \delta'(k^2)$. The two distributions have overlapping singular support, thus their product is not well-defined. In order to make it well-defined, one would have to add a nonlocal counterterm, i.e., there are nonlocal renormalization ambiguities (see Section~\ref{sec:infrared} for details). Then the theory loses its predictive power.

A way to circumvent this problem might be to look at a supersymmetric version of the model. It has been noticed \cite{Matusis}, that the term~(\ref{eq:Hayakawa_Sigma_2}) disappears in such a setting when the modified Feynman rules are employed.
We will show that this is also true in the Yang-Feldman approach.

In gauge theories on noncommutative spaces it is generally a subtle point to construct local gauge invariant quantities (observables). In the approach via the Seiberg-Witten map, one can use the field strength $\tilde F_{\mu \nu}$ corresponding to the ordinary gauge fields $\tilde A_\mu$ and smear it with some test function. In the unexpanded case, the situation is more difficult, the reason being that the $q^\mu$'s do not transform covariantly, so that multiplication with a (test) function $f(q)$ is no homomorphism. The only way to construct local observables that is known so far is to use the covariant coordinates~\cite{Madore}
\begin{equation}
\label{eq:CovCoor}
 X^\mu = q^\mu + e \Theta^{\mu \nu} A_\nu.
\end{equation}
They transform covariantly, so $\int \ud^4 q \  f(X) F_{\mu \nu}$ is gauge invariant, due to the cyclicity of the integral. In the context of classical electrodynamics on the noncommutative Minkowski space, the covariant coordinates have been used in~\cite{NCED} to compute the effect of a constant background field on the propagation of light. Here we are going to use them in the quantized setting. We note that most of the computations done in the literature in the unexpanded approach did not take the covariant coordinates into account and thus were not considering gauge invariant quantities. To the best of our knowledge, the only exceptions are \cite{Gross, Rozali}, who, however, do not consider all possible contractions of the photon fields stemming from the covariant coordinates (only contractions where both fields stem from the same observable).

For our study of the supersymmetric version of the model, we would like to have observables that are not only invariant under gauge, but also under supersymmetry transformations. We will show that this can be achieved by considering a supersymmetric version of the covariant coordinates.

In the present paper, we compute the two-point function of the interacting field strength at second order in $e$. We consider pure electrodynamics, i.e., we do not include fermionic matter fields\footnote{This can be justified by the fact that at $\order(e^2)$, the fermion contributions are exactly as in the commutative case~\cite{hayakawa} and thus not of interest for our study.}. As can be anticipated from the presence of the coupling constant in (\ref{eq:CovCoor}), the effect of the covariant coordinates can be accounted for perturbatively. One may hope that they help to tame the bad infrared behavior indicated in~(\ref{eq:SelfEnergy}). However, we find that
\begin{enumerate}
\item
\label{enum:1}
In order to set up the free theory, we have to use 
a normal ordered version of functions of the covariant coordinates. This could also be interpreted as the subtraction of nonlocal counterterms.
\item 
\label{enum:2}
From the terms in which one power of $e$ stems from the interaction and one from the covariant coordinates we get a contribution that is nonlocal and divergent (a nonlocal expression multiplied with a divergent quantity).
\item 
\label{enum:3}
The terms where the two powers of $e$ solely stem from the interaction give essentially the same result as the modified Feynman rules, i.e., the nonplanar contributions to the singular part of the spectrum are of the form~(\ref{eq:SelfEnergy}) with $\Sigma_{1/2}$ as in~(\ref{eq:Hayakawa_Sigma_1}, \ref{eq:Hayakawa_Sigma_2}).
\end{enumerate}
Since nonlocal counterterms seem to be unavoidable, the theory can at best be considered as effective.

When we consider the supersymmetric version of the model, we find that the nonlocal divergences described in \ref{enum:2} and \ref{enum:3} in the above list are cancelled. As mentioned above, the cancellation of \ref{enum:3} was already found in the context of the modified Feynman rules \cite{Matusis}. The cancellation of \ref{enum:2} is only possible because a supersymmetric version of the covariant coordinates is employed. The problematic term mentioned in \ref{enum:1} remains, but it can also be interpreted in a natural way as a redefinition of functions of the covariant coordinates. Apart from this term, the only modification of the two-point function is then a momentum-dependent field strength normalization. It can be interpreted as giving rise to acausal effects. Interestingly, these are independent of the noncommutativity scale (in particular, they are not restricted to this scale).

The paper is organized as follows. In the next section we shortly discuss calculus on the noncommutative Minkowski space and introduce the Lagrangean, including a gauge fixing via the BRST formalism. In Section~\ref{sec:YF} we set up the Yang-Feldman series. It is argued that one should replace the usual product of quantum fields by a symmetrized product. We then derive graphical rules for the model in Section~\ref{sec:GraphicalRules}. In Section~\ref{sec:CovCoor} we introduce the quantized covariant coordinates. In Sections~\ref{sec:2pt}--\ref{sec:Full2pt} we compute the two-point function of the interacting field strength. In Section~\ref{sec:infrared} we discuss the well-definedness of products of distributions appearing in the two-point function and the need for nonlocal counterterms. In most of these sections, there is a subsection that deals with the modifications that are necessary to accommodate supersymmetry. The results of these subsections are combined in Section~\ref{sec:SUSY}, where the two-point function in the supersymmetric case is computed. Its implication for causality is discussed in Section~\ref{sec:NonlocalEffects}. We conclude in Section~\ref{sec:Conclusion}.

The results presented here stem from the PhD thesis \cite{diss} of the author, which was written under the supervision of Klaus Fredenhagen at the II.~Institut f\"ur Theoretische Physik at the Universit\"at Hamburg.

\section{Setup}

Functions (including classical fields) on the noncommutative Minkowski space are defined \`a la Weyl: Let $f$ be a function on the ordinary Minkowski space and $\hat f$ its Fourier transform. Then we define
\begin{equation}
\label{eq:Weyl}
f(q) = (2\pi)^{-2} \int \ud^4 k \ \hat{f}(-k) e^{ikq}.
\end{equation}
One assumes that the commutation relations \eqref{eq:q_comm} can be integrated to yield
\[
e^{ikq} e^{ipq} = e^{i(k+q)q} e^{-\frac{i}{2} k \Theta p}.
\]
On the exponentials $e^{ikq}$, one defines
\[
\int \ud^4q \ e^{ikq} = (2\pi)^4 \delta(k).
\]
It is easy to see that this map is cyclic and reproduces the integral on the ordinary Minkowski space when applied to $f(q)$. Derivatives can be defined via translations,
\[
\del_\mu f(q) = \frac{\ud}{\ud t} f(q + t e^\mu),
\]
or, using the commutation relations, by
\begin{equation}
\label{eq:CommutationDerivative}
\del_\mu f(q) = - i \Theta^{-1}_{\mu \nu} [q^\nu, f(q)].
\end{equation}

We want to quantize noncommutative electrodynamics via the Yang-Feldman formalism. For this we need a well-posed Cauchy problem and thus have to break gauge invariance. We use the BRST formalism and introduce ghosts and antighosts $c$ and $\bar{c}$ and the Nakanishi-Lautrup field $B$. Our Lagrangean is then
\begin{equation}
\label{eq:Lagrangean}
 L = - \tfrac{1}{4} F_{\mu \nu} F^{\mu \nu} + \del_{\mu} B A^{\mu} + \tfrac{\alpha}{2} B^2 -  \del_{\mu} \bar{c} D^{\mu} c .
\end{equation}
Here we used the notation $D_{\mu} c = \del_{\mu} c - i e [A_{\mu}, c]$. The corresponding action is invariant under the BRST transformation
\begin{align*}
 \delta_\xi A_{\mu} & = \xi D_{\mu} c, \\
 \delta_\xi c & = \xi \tfrac{i}{2} e \{c,c\}, \\
 \delta_\xi \bar{c} & = \xi B, \\
 \delta_\xi B & = 0,
\end{align*}
where $\xi$ is an infinitesimal anticommuting parameter. It is straightforward to show that the first term on the \rhs of (\ref{eq:Lagrangean}) transforms covariantly under $\delta_\xi$,
\begin{equation*}
 F_{\mu \nu} F^{\mu \nu} \to - i e \xi [F_{\mu \nu} F^{\mu \nu},c],
\end{equation*}
and the remaining terms are invariant under $\delta_\xi$. Thus, by cyclicity, the action obtained by integrating over the Langrangean is invariant. Furthermore, if we write $\delta_\xi = \xi \delta'$, then $\delta'$ is nilpotent, as usual.

From the above Lagrangean, we get the equations of motion
\begin{subequations}
\begin{align}
\label{eq:eom1}
 \Box A^{\mu} - \del^{\mu} \del_{\nu} A^{\nu} + \del^{\mu} B & = i e \del_{\nu} [A^{\nu}, A^{\mu}] + i e [A_{\nu}, F^{\nu \mu}] + i e \{ \del^{\mu} \bar{c}, c\}, \\
\label{eq:eom2}
 \alpha B - \del_{\mu} A^{\mu} & = 0, \\
\label{eq:eom3}
 \Box c & = i e \del_{\mu} [A^{\mu}, c], \\
\label{eq:eom4}
 \Box \bar{c} & = i e [A^{\mu}, \del_{\mu} \bar{c}].
\end{align}
\end{subequations}

In \cite{diss}, it is shown that the corresponding BRST (and ghost) current is not conserved. This is a phenomenon that often occurs in interacting NCFTs \cite{Conservation, NCDispRel}.
That the interacting BRST current is not conserved is not problematic, as long as the interacting BRST operator $\delta'_{\text{int}}$ is still nilpotent (and thus the physical state space can be defined in the usual way). This is the case if the renormalized Lagrangean is still of the appropriate form, i.e., if the usual relations between counterterms hold. That this is the case at the one-loop level has been shown in the setting of the modified Feynman rules in~\cite{Martin}. There, only the local counterterms, i.e., those that arise from planar graphs, were taken into account. As we show below, these are the same in the Yang-Feldman formalism. But we will argue in the following that nonlocal counterterms are necessary. Then the situation becomes more involved. In any case $\delta'$ is nilpotent at tree level, and thus the Ward identity will be satisfied at tree level in any quantization scheme that respects the classical equations of motion,
as the Yang-Feldman formalism. In Appendix~\ref{app:Ward}, we explicitly show this for Compton scattering at second order, a process which is known to violate the Ward identity at tree level in the Hamiltonian approach to NCQED \cite{Ohl}. 

\subsection{The supersymmetric case}
A simple way to introduce supersymmetry is to add a Weyl fermion~$\lambda$, the photino, and an auxiliary field $D$. Both fields transform in the adjoint representation. Then one adds the terms
\begin{equation}
\label{eq:SNCQED_Action}
 i \bar \lambda \bar \sigma^\mu D_\mu \lambda + 2 D^2
\end{equation}
to the Lagrangean \eqref{eq:Lagrangean}. Properly, one should also add superpartners of the ghosts, but these do not contribute to the two-point function at second order, so we ignore them here.
In order to construct observables that are invariant under supersymmetry transformations (cf. Section~\ref{sec:CovCoor}), it is advantageous to work in the superfield formalism. Thus, we embed our fields in the vector multiplet $V$ by defining\footnote{We assume that the anticommuting coordinates commute with the $q^\mu$'s. For our conventions on spinors and supersymmetry, we refer to Appendix~\ref{app:SUSY}.}
\begin{equation*}
 V = - \theta \sigma^\mu \bar \theta A_\mu + i \theta^2 \bar \theta \bar{\lambda} - i \bar{\theta}^2 \theta \lambda + \theta^2 \bar{\theta}^2 D.
\end{equation*}
Because of the anticommutativity of the $\theta$'s, we have $V^3=0$. An infinitesimal gauge transformation can now be written as
\begin{equation}
\label{eq:delta_V}
 \delta_\Lambda V = \tfrac{i}{2e} (\bar{\Lambda} - \Lambda) - \tfrac{i}{2} [V, \bar{\Lambda}+ \Lambda],
\end{equation}
where $\Lambda$ is a chiral field given by
\begin{equation*}
 \Lambda(q, \theta, \bar \theta) = e^{-i \theta \sigma^\mu \bar \theta \del_\mu} \chi(q).
\end{equation*}
Here $\chi$ is the usual infinitesimal gauge parameter. Because of the rather complicated form of the gauge transformation~(\ref{eq:delta_V}), it is advantageous to introduce yet another superfield, namely
\begin{equation*}
 W_\alpha = - \tfrac{1}{4e} \bar{D}^2 (e^{-2eV} D_\alpha e^{2eV}) = - \tfrac{1}{2} \bar{D}^2 ( D_\alpha V - e [V, D_\alpha V] ).
\end{equation*}
It transforms in the adjoint representation, i.e., as
\begin{equation}
\label{eq:delta_W}
 \delta_\Lambda W_\alpha = i [\Lambda, W_\alpha].
\end{equation}
Because of the anticommutativity of the $\bar D^\dotalpha$'s, $W_\alpha$ is chiral. In component form, it is given by
\begin{equation}
\label{eq:W_alpha_Components}
 W_\alpha = - 2 i \lambda_\alpha + 2 i {{\sigma^{\mu \nu}}_\alpha}^\beta \theta_\beta F_{\mu \nu} + 4 \theta_\alpha D - 2 \theta^2 \sigma^\mu_{\alpha \dotalpha} D_\mu \bar \lambda^\dotalpha + \order(\bar \theta).
\end{equation}
The action, without the ghost part, can then be expressed as
\begin{equation*}
 S = \tfrac{1}{16} \int \ud^6q \ W^\alpha W_\alpha + \text{ h.c.}
\end{equation*}
Here the integral over $\ud^6q$ denotes the integral over $\ud^4q$ of the $\theta^2$ component.

\section{The Yang-Feldman procedure}
\label{sec:YF}

Using the equations of motion from the previous section, we can now set up the Yang-Feldman series.
Choosing Feynman gauge, $\alpha =1$, and using~(\ref{eq:eom2}), we can eliminate the $B$-field from~(\ref{eq:eom1}):
\begin{equation}
\label{eq:eom1b}
 \Box A^{\mu} = i e \del_{\nu} [A^{\nu}, A^{\mu}] + i e [A_{\nu}, F^{\nu \mu}] + i e \{ \del^{\mu} \bar{c}, c \}.
\end{equation}
The interacting fields $A^\mu$, $c$ and $\bar{c}$ are now defined as a formal power series in the coupling constant $e$, i.e.,
\[
 A^\mu = \sum_{n=0}^\infty e^n A_n^\mu,
\]
and analogously for $c$ and $\bar c$. Plugging this ansatz in \eqref{eq:eom1b}, one finds that $A_0$, $c_0$, and $\bar{c}_0$ satisfy the free field equation. In the Yang-Feldman approach, they are identified with the incoming field. Thus, the higher order components are obtained by convolution with the retarded propagator. Using (\ref{eq:eom1b}), (\ref{eq:eom3}) and (\ref{eq:eom4}), one thus obtains
\begin{align*}
 A^{\mu}_i & = i \int \ud^4x \ \Delta_R(x) \Big\{ - i  \sum_{j+k+l=i-2} \left[A_{\nu j}, \left[ A^{\nu}_k , A^{\mu}_l \right] \right] \\
 & \qquad \qquad + \sum_{j+k=i-1} \left( \del_{\nu} \left[A^{\nu}_j, A^{\mu}_k \right]  + \left[ A_{\nu j}, \del^{\nu} A^{\mu}_k - \del^{\mu} A^{\nu}_k \right] + \left\{ \del^{\mu} \bar{c}_j, c_k \right\} \right) \Big\}_x \\
 c_i & = i \int \ud^4 x \ \Delta_R(x) \Big\{ \sum_{j+k=i-1} \del_{\mu} \left[ A^{\mu}_j, c_k \right] \Big\}_x \\
 \bar{c}_i & = i \int \ud^4 x \ \Delta_R(x) \Big\{ \sum_{j+k=i-1} \left[ A^{\mu}_j, \del_{\mu}  \bar{c}_k \right] \Big\}_x
\end{align*}
Here we used a notation where the subscript $x$ stands for translation along $x$, i.e., $f(q)_x = f(q-x)$. Thus, at first order, the interacting fields are
\begin{subequations}
\label{eq:X1}
\begin{align}
\label{eq:A1}
 A_1^{\mu} & = i \int \ud^4x \ \Delta_R(x) \left\{ \del_{\lambda} [A_0^{\lambda}, A_0^{\mu}] + [A_0^{\lambda}, \del_{\lambda} A_0^{\mu}] - [A_0^{\lambda}, \del^{\mu} A_{0 \lambda}] + \{ \del^{\mu} \bar{c}_0, c_0 \} \right\}_x \\ 
\label{eq:c1} 
 c_1 & = i \int \ud^4x \ \Delta_R(x) \left\{ \del_{\mu} [A_0^{\mu}, c_0] \right\}_x \\
\label{eq:c1_bar} 
 \bar{c}_1 & = i \int \ud^4x \ \Delta_R(x) \left\{ [A_0^{\mu}, \del_{\mu} \bar{c}_0] \right\}_x.
\end{align}
\end{subequations}
The photon field at second order is then
\begin{subequations}
\label{eq:A2}
\begin{align}
\label{subeq:A2_1}
 A_2^{\mu} = i \int \ud^4x \ \Delta_R(x) \Big\{ & \del_{\lambda} [A_1^{\lambda}, A_0^{\mu}] + \del_{\lambda} [A_0^{\lambda}, A_1^{\mu}] + [A_1^{\lambda}, \del_{\lambda} A_0^{\mu}] \\
\label{subeq:A2_2}
 & + [A_0^{\lambda}, \del_{\lambda} A_1^{\mu}] - [A_1^{\lambda}, \del^{\mu} A_{0 \lambda}] - [A_0^{\lambda}, \del^{\mu} A_{1 \lambda}] \\
\label{subeq:A2_3} 
 & + \{ \del^{\mu} \bar{c}_1, c_0 \} + \{ \del^{\mu} \bar{c}_0, c_1 \} \\
\label{subeq:A2_4}  
 & - i [A_{0 \lambda}, [A_0^{\lambda}, A_0^{\mu}]] \Big\}_x 
\end{align}
\end{subequations}

As proposed in~\cite{dfr}, the quantized free fields are elements of (or rather affiliated to) $\mathfrak{F} \otimes \mathcal{E}_\Theta$, where $\mathfrak{F}$ is the algebra of operators on Fock space and $\mathcal{E}_\Theta$ is the algebra generated by the quantum coordinates $q^{\mu}$. Explicitly, we have
\begin{equation*}
 A^{\mu}(q) = (2\pi)^{-2} \int \ud^4k \ \hat{A}^{\mu}(k) \otimes e^{-ikq} 
\end{equation*}
with
\begin{equation*}
 \hat{A}^{\mu}(k) = (2\pi)^{\frac{1}{2}} \delta(k^2) \left( \theta(k_0) a^{\mu}(k) + \theta(-k_0) a^{\mu}(-k)^* \right),
\end{equation*}
and analogously for $c(q)$ and $\bar{c}(q)$. The operators $a^{\mu}(k)$ and their adjoints fulfill the usual commutation relations.
In order to give some meaning to the \rhs of \eqref{eq:X1} and \eqref{eq:A2}, we have to specify how products of quantum fields are defined. In the following, we use the product
\begin{equation}
\label{eq:SymmProd}
 ( \phi \otimes f ) \cdot ( \psi \otimes g ) = \tfrac{1}{2} (\phi \psi + \psi \phi) \otimes f g,
\end{equation}
since it leads the correct commutative limit of commutator terms (which are the relevant one here). For a detailed discussion of the choice of this product, we refer to \cite[Section~6.3]{diss}.

\subsection{Dispersion relations in the Yang-Feldman formalism}
\label{sec:DispRel}
We briefly discuss how dispersion relations are computed in the Yang-Feldman formalism. This also serves as an illustration for how to perform renormalization in this framework. For simplicity, we restrict to the case of a scalar field and to the one-loop level, i.e., to $n=2$ in the $\phi^3$ case and $n=1$ in the $\phi^4$ case. We then write the corresponding component of the interacting field in the form
\begin{equation}
\label{eq:ScalarSelfEnergy}
 \hat{\phi}_n(k) = (2\pi)^2 \hat{\Delta}_{R}(k) \Sigma(k) \hat{\phi}_0(k) + \text{c.s.}
\end{equation}
where ``c.s.'' stands for the continuous part of the spectrum. Here $\Sigma(k)$ is the self-energy. In the noncommutative case it is a function of $k^2$ and $(k \Theta)^2$. The above equation has the form of a mass renormalization. Possible infinities can be absorbed by adding a mass counterterm $\lambda^n \delta m^2 \phi^2$ to the Lagrangean.
Since by the basic principles of renormalization theory such counterterms are local, the influence of the $(k \Theta)^2$ dependence of the self-energy can not be absorbed. This leads to a modified dispersion relation
\[
  k^2 - m^2 + \lambda^n \Sigma_{\text{ren}}(m^2, (\Theta k)^2) = 0,
\]
where $\Sigma_{\text{ren}}$ is the renormalized self-energy. In the case of the $\phi^3$ model the resulting distortion is moderate \cite{NCDispRel}, while it is quite strong in the $\phi^4$ model \cite{Quasiplanar}. In the case of NCQED, one also expects a severe distortion, similar to what was found in the setting of the modified Feynman rules \cite{Matusis}.

However, we may also take a different viewpoint. In \eqref{eq:ScalarSelfEnergy} we have a product of the distributions $\hat{\Delta}_R(k)$, $\Sigma(k)$ and $\hat{\phi}_0(k)$ (to be precise the latter is a distribution valued operator). As shown in \cite{AdLim, NCDispRel}, the product of $\hat{\Delta}_R(k)$ and $\hat{\phi}_0(k)$ can be defined rigorously in the sense of a weak adiabatic limit, i.e., the two-point function of such an interacting field is well-defined. But singularities in $\Sigma(k)$ might spoil this. Formally,  \eqref{eq:ScalarSelfEnergy} leads to the following term in the two-point function of the interacting field:
\[
 \bra{\Omega} \left( \hat{\phi}_n(k) \hat{\phi}_0(p) + \hat{\phi}_0(k) \hat{\phi}_n(p) \right) \ket{\Omega} = - (2\pi)^2 \delta(k+p) \Sigma(k) \tfrac{\del}{\del m^2} \hat{\Delta}_+(k).
\]
We will see in Section~\ref{sec:infrared} that in the case of NCQED the product of the self-energy and the derivative of $\hat{\Delta}_+$ is not well-defined on a one-dimensional submanifold of $\R^4$. Fixing the ambiguity in the definition of this product thus introduces a continuum of renormalization conditions (one for each point on this line). Thus, the theory can at best be considered as effective.

\subsection{The supersymmetric case}
\label{sec:YFSUSY}
From the term~(\ref{eq:SNCQED_Action}), we obtain the following equations of motion for the photino:
\begin{equation*}
 i \bar \sigma^\mu \del_\mu \lambda + e \bar \sigma^\mu [A_\mu, \lambda] = 0, \quad
 -i \sigma^\mu \del_\mu \bar \lambda - e \sigma^\mu [A_\mu, \bar \lambda] = 0.
\end{equation*}
Thus, the first order contribution to the interacting field is
\begin{equation*}
 \lambda_1 = \int \ud^4x \ S_R(x) \left\{ - \bar \sigma^\mu [A_{\mu 0}, \lambda_0] \right\}_x, \quad
 \bar \lambda_1 = \int \ud^4x \ \bar S_R(x) \left\{ \sigma^\mu [A_{\mu 0}, \bar \lambda_0] \right\}_x.
\end{equation*}
Here we used the propagators
\begin{equation*}
 S_R = - i \sigma^\nu \del_\nu \Delta_R, \quad
 \bar S_R = i \bar \sigma^\nu \del_\nu \Delta_R.
\end{equation*}

The introduction of the photino changes the equation of motion for $A^\mu$. Instead of \eqref{eq:eom1b}, it is now given by
\begin{equation}
\label{eq:ExtraTerm}
 \Box A^{\mu} = i e \del_{\nu} [A^{\nu}, A^{\mu}] + i e [A_{\nu}, F^{\nu \mu}] + i e \{ \del^{\mu} \bar{c}, c \} + e \sigma^\mu_{\alpha \dot \alpha} \{ \lambda^\alpha, \bar \lambda^{\dot \alpha} \}.
\end{equation}

\section{Graphical rules}
\label{sec:GraphicalRules}

In order to efficiently deal with the different terms in the Yang-Feldman series, we introduce a set of graphical rules. In these, a double line stands for a retarded propagator, a vertex for a product of fields, and an open circle for a free field. Thus, $A_1^\mu$ can be represented by
\begin{center}
\includegraphics{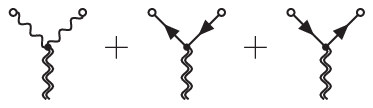}
\end{center}
and similarly for $c_1$ and $\bar{c}_1$.
Here an upward pointing straight line stands for an antighost $\bar c$ and a downward pointing one for a ghost $c$ emanating from the vertex. In the process of recursively constructing the interacting field, the free field, i.e., the open circle and the single line, may be replaced by a corresponding building block of higher order. The terms \eqref{subeq:A2_1} and \eqref{subeq:A2_2}, for example, are represented by the graphs
\begin{center}
\includegraphics{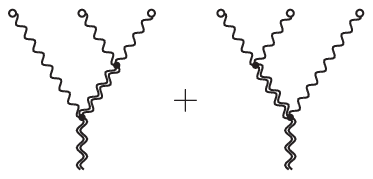}
\end{center}
Thus, apart from the cubic vertices stemming from $c_1$ and $\bar{c}_1$, the only missing building block is the quartic photon vertex from \eqref{subeq:A2_4}:
\begin{center}
\includegraphics{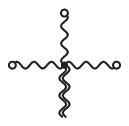}
\end{center}

Now we want to state the graphical rules.
We begin with the computation of the cubic photon vertex
\begin{center}
\includegraphics{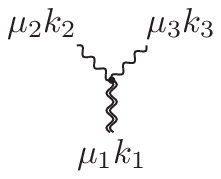}
\end{center}
where the $k_i$ are incoming momenta.
The vertex factor is obtained from the part of $\hat A^{\mu_1}_1(k_1)$ that is quadratic in $A_0$, i.e., the first three terms in (\ref{eq:A1}):
\begin{align*}
\hat{A}^{\mu_1}_1(k_1) & = - (2\pi)^{-4} \tfrac{1}{2} \hat{\Delta}_R(k_1) \int \ud^4p_1 \ud^4p_2 \ \int \ud^4q \ [e^{ip_1q}, e^{ip_2q}] e^{ik_1q} \\ 
& \quad \times \left\{ (p_1+2p_2)_{\rho} \left( \hat{A}_0^{\rho}(-p_1) \hat{A}_0^{\mu_1}(-p_2) + \hat{A}_0^{\mu_1}(-p_2) \hat{A}_0^{\rho}(-p_1) \right)  \right. \\
 & \quad \quad \left. - p_2^{\mu_1} \left( \hat{A}_0^{\rho}(-p_1) \hat{A}_{0 \rho}(-p_2) + \hat{A}_{0 \rho}(-p_2) \hat{A}_0^{\rho}(-p_1) \right) \right\}. 
\end{align*}
Now one has to equate the momentum and Lorentz index of the left/right field operators with those of the lines leaving the above vertex to the left/right. For the first term, this means replacing $p_1$ by $k_2$, $\rho$ by $\mu_2$, $p_2$ by $k_3$ and multiplying by $g^{\mu_1 \mu_3}$. This way, we find the vertex factor
\begin{equation}
\label{eq:CubicVertex}
 i e \left( g^{\mu_1 \mu_2} (k_1-k_2)^{\mu_3} + g^{\mu_3 \mu_1} (k_3-k_1)^{\mu_2} + g^{\mu_2 \mu_3} (k_2-k_3)^{\mu_1} \right) \sin \tfrac{k_2 \Theta k_3}{2} \delta(\sum k_i).
\end{equation}
We remark that it is invariant under permutations of the legs.

In order to find the factor for the quartic vertex
\begin{center}
\includegraphics{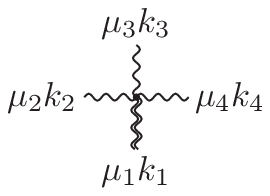}
%
%
%
\end{center}
one has to consider the term (\ref{subeq:A2_4}). We find the vertex factor
\begin{multline*}
 (2\pi)^{-2} e^2 \delta(\sum k_i) \left\{ \sin \tfrac{k_1 \Theta k_2}{2} \sin \tfrac{k_3 \Theta k_4}{2} \left( g^{\mu_1 \mu_3} g^{\mu_2 \mu_4} - g^{\mu_1 \mu_4} g^{\mu_2 \mu_3} \right) \right. \\
 \left. + \sin \tfrac{k_2 \Theta k_3}{2} \sin \tfrac{k_4 \Theta k_1}{2}
 \left( g^{\mu_1 \mu_3} g^{\mu_2 \mu_4} - g^{\mu_1 \mu_2} g^{\mu_3 \mu_4} \right) \right\}.
\end{multline*}
Note that this is not symmetric under permutations of the legs and does not coincide with the vertex factor found in the setting of the modified Feynman rules \cite{hayakawa}. The reason is that there a total symmetrization of the multiple products of fields is implicitly assumed, cf. \cite[Remark 6.3.2]{diss}.

It remains to treat the ghost vertices. From the fourth term in (\ref{eq:A1}), we obtain
\begin{center}
\includegraphics{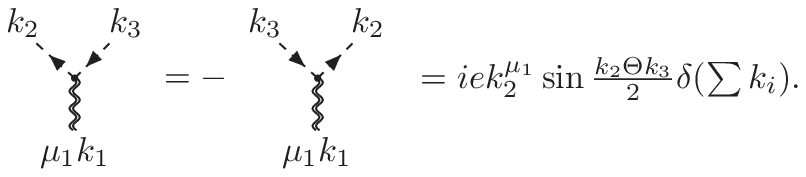}
\end{center}
Note that we antisymmetrized the ghost field operators.
Here the upward pointing line stands for an antighost $\bar c$ and the downward pointing line for a ghost $c$ emanating from the vertex.
For the vertices with incoming ghost or antighost, we find, using (\ref{eq:c1}) and (\ref{eq:c1_bar}),
\begin{center}
\includegraphics{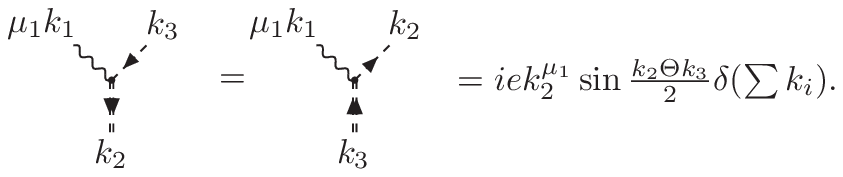}
\end{center}
The factors for the vertices where the photon leaves to the right (with the same momentum) are the same, because photon and ghost fields commute.

It remains to define the propagators. As already mentioned, double lines stand for a retarded propagator. Contracting two free fields, i.e., joining two open circles, yields the two-point function. Thus, the photon propagators are given by
\begin{center}
\includegraphics{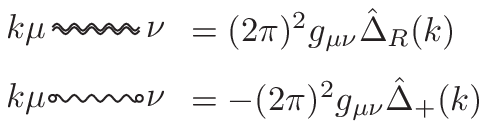}
\end{center}
and the ghost propagators by 
\begin{center} 
\includegraphics{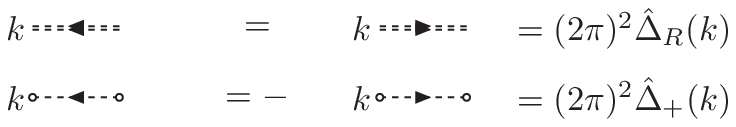}
\end{center}

\subsection{The supersymmetric case}
For the photino, we obtain, in the same manner, the vertex factors (cf. Section~\ref{sec:YFSUSY})
\begin{center}
\includegraphics{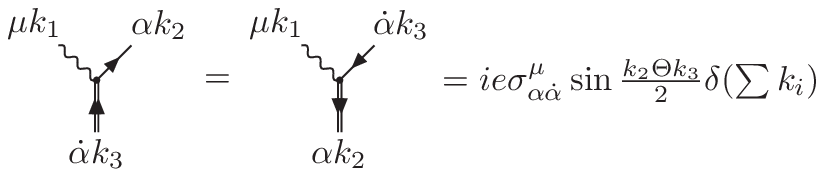}
\end{center}
Here the upward pointing line stands for the photino $\lambda$ and the downward pointing line for the anti-photino $\bar \lambda$. The photino propagators are given by 
\begin{center}
\includegraphics{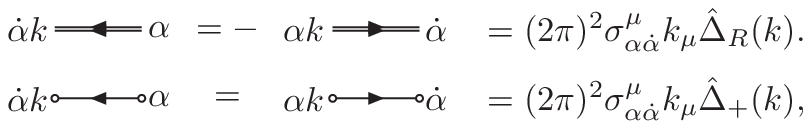}
\end{center}
For the retarded propagator, the direction from left to right depicted here corresponds to the upward direction in graphs.
The last term on the \rhs of \eqref{eq:ExtraTerm} leads to the vertex factors
\begin{center}
\includegraphics{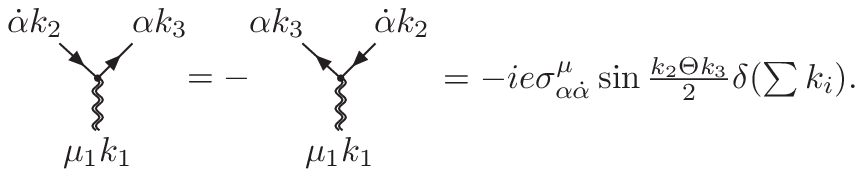}
\end{center}

\section{Covariant Coordinates}
\label{sec:CovCoor}

In this section we introduce the covariant coordinates. We show how to quantize them and derive some useful tools for later calculations. We recall that they are given by
\begin{equation*}
 X^{\mu} = q^{\mu} + e \Theta^{\mu \nu} A_{\nu}.
\end{equation*}
Using \eqref{eq:CommutationDerivative}, it is straightforward to show that they transform covariantly under a BRST transformation:
\begin{equation*}
 \delta_\xi X^\mu = - i e \xi [X^\mu, c].
\end{equation*}
From the cyclicity of the integral it follows that
\begin{equation}
\label{eq:NCQEDObservable}
 \int \ud^4 q \ f^{\mu \nu}(X) F_{\mu \nu}
\end{equation}
is BRST invariant and thus an observable.
\begin{remark}
We comment on the relation of this observable to the expression $\int \ud^4 q \ f^\mu A_\mu$ usually used to compute the photon two-point function. Using~(\ref{eq:F}), one can bring~(\ref{eq:NCQEDObservable}) to this form by setting $f^\mu = D_\nu f^{\mu \nu}(X) + \del_\nu f^{\mu \nu}(X)$. It is easy to see that this $f^\mu$ is neither (covariantly) conserved, nor does it transform covariantly. We also remark that the use of covariant coordinates can not be evaded by focussing on matter observables and inferring the dispersion relation of the photon by its effect on these. The reason is that the current transforms covariantly, so it should also be evaluated in covariant coordinates.
\end{remark}
As for the $q's$, we define functions of the covariant coordinates \`a la Weyl, i.e.,
\begin{equation*}
 f(X) = (2\pi)^{-2} \int \ud^4k \ \hat{f}(-k) e^{ikX}.
\end{equation*}
In order to make sense of the Weyl factor $e^{ikX}$, we express it as a formal power series in the coupling constant $e$. This is logically consistent, since the field strength $F_{\mu \nu}$ in (\ref{eq:NCQEDObservable}) is also given as a formal power series in $e$, so that (\ref{eq:NCQEDObservable}) is gauge invariant at each order of $e$. The $N$th order part of $e^{ikX}$ can be found by computing\footnote{Note that we are not using the symmetrized product~(\ref{eq:SymmProd}) for the definition of $e^{ikX}$, since it is not clear in which order one should do that (we recall that the symmetrized product is not associative). However, we use it for the product of $f^{\mu \nu}(X)$ and~$F_{\mu \nu}$.}
\begin{equation*}
 \frac{1}{N!} \left( \frac{\ud}{\ud \lambda} \right)^N e^{C+\lambda D} |_{\lambda=0} = \sum_{n_0, \dots , n_{N}}  \frac{C^{n_0} D C^{n_1}  \dots D C^{n_N}}{(n_0 + \dots + n_N+N)!}
\end{equation*}
Here $C$ and $D$ stand for arbitrary elements of some algebra. We are of course interested in the case where $C$ is replaced by $ikq$ and $D$ by $ik_{\mu} \Theta^{\mu \nu} A_{\nu}$. For the $i$th such $A$, we write 
\[
A_{\nu_i}(q) = (2\pi)^{-2} \int \ud p_i \ \hat{A}_{\nu_i}(p_i) e^{-ip_i q}.
\]
The $\hat{A}_{\nu_i}(p_i)$'s can be pulled out of the expression. If we then commute all the $e^{-ip_iq}$'s to the right, we will have to replace $(ikq)^{n_i}$ by $(i k(q - \Theta \sum_{j \leq i} p_j))^{n_i}$. In order to deal with the resulting expression, we need the following
\begin{lemma}
\label{lemma}
Let $x, y_i$ be pairwise commuting elements of an algebra. Then
\begin{equation*}
 \sum_{n_0, \dots n_N} \frac{x^{n_0} (x+y_1)^{n_1} \dots (x+y_N)^{n_N}}{(n_0 + \dots + n_N+N)!} = e^x \sum_{n_1, \dots n_N} \frac{y_1^{n_1} \dots y_N^{n_N}}{\left( \sum_{i=1}^N (n_i+1) \right)!}.
\end{equation*}
\end{lemma} 

The proof can be found in Appendix~\ref{app:lemma}. We define
\begin{equation*}
 P_N(y_1, \dots, y_N) = \sum_{n_1, \dots , n_N=0}^\infty \frac{\prod_{m=1}^N \left( \sum_{n=1}^m y_n \right)^{n_m}}{\left( \sum_{i=1}^N (n_i+1) \right)!}.
\end{equation*}
Thus, one obtains
\begin{multline}
\label{eq:e_ikX}
 e^{ikX} = e^{ikq} \sum_{N=0}^\infty (ie)^N (2\pi)^{-2N} \int \prod_{i=1}^N \ud^4k_i \ e^{-i k_1q} \dots e^{-ik_Nq} \\
 \times k \Theta \hat A(k_1) \dots k \Theta \hat A(k_N) P_N(-ik\Theta k_1, \dots, -ik\Theta k_N).
\end{multline}
In Section~\ref{sec:FreePart} we will show that the products of fields in this expression are not well-defined and require normal ordering.

\begin{remark}
\label{rem:Equivalence}
Equation~(\ref{eq:e_ikX}) is equivalent to the formula, cf.~\cite{BakLeePark},
\begin{equation*}
 e_\star^{ikX} = e^{ikx} \star \bar{\mathrm{P}}_\star e^{i e \int_0^1 \ud t \ k \sigma A(x+t k \Theta)}.
\end{equation*}
Here $\bar{\mathrm{P}}_\star$ is the anti path ordered $\star$-product. This is proven in \cite[App. B.6]{diss}.
\end{remark}

\subsection{The supersymmetric case}
When discussing the supersymmetric version of the model, one immediately recognizes that while the observable $\int \ud^4q \ f^{\mu \nu}(X) F_{\mu \nu}$ is gauge invariant, it is not invariant under the supersymmetry transformation\footnote{Note that this transformation is nonlinear. This is due to the fact that we implicitly adopted the Wess--Zumino gauge.}
\begin{align*}
 s_\varepsilon A^\mu & = i \varepsilon \sigma^\mu \bar \lambda + i \bar \varepsilon \bar \sigma^\mu \lambda, \\
 s_\varepsilon \lambda_\alpha & = - (\sigma^{\mu \nu})_\alpha^{\ \beta} F_{\mu \nu} \varepsilon_\beta - 2 i \varepsilon_\alpha D, \\
 s_\varepsilon D & = \tfrac{1}{2} ( \bar \varepsilon \bar \sigma^\mu D_\mu \lambda - \varepsilon \sigma^\mu D_\mu \bar \lambda ).
\end{align*}
Here $\varepsilon$ is an infinitesimal anticommuting parameter. The easiest way to construct observables for the field strength that are not only invariant under gauge, but also under supersymmetry transformations is to express
\begin{equation*}
 \int \ud^4q \ f^{\mu \nu}(q) F_{\mu \nu}
\end{equation*}
in superfield form. Using the component form (\ref{eq:W_alpha_Components}) of $W_\alpha$, one can show that with
\begin{equation}
\label{eq:f_alpha}
 f^\alpha = \tfrac{i}{2} \theta^\beta {{\sigma^{\mu \nu}}_\beta}^\alpha  f_{\mu \nu},
\end{equation}
one obtains
\begin{equation*}
 \int \ud^4q \ f^{\mu \nu}(q) F_{\mu \nu} = \int \ud^6q \ f^\alpha(q) W_\alpha + \text{ h.c.}
\end{equation*}
It remains to find the appropriate covariant coordinates. For this, we make the ansatz
\begin{equation*}
 X^\mu = q^\mu + e \Theta^{\mu \nu} Y_\nu.
\end{equation*}
In order for $X^\mu$ to transform covariantly, $Y_\nu$ must transform as
\begin{equation}
\label{eq:delta_Y}
 \delta_\Lambda Y_\nu = \tfrac{1}{e} \del_\nu \Lambda + i [\Lambda, Y_\nu],
\end{equation}
cf. \eqref{eq:delta_W}. We also know that the body of $Y_\nu$ should contain the vector potential $A_\nu$. It is then not so difficult to guess
\begin{equation*}
 Y_\nu = \tfrac{1}{4e} \bar{\sigma}_\nu^{\dot \alpha \alpha} \bar{D}_{\dot \alpha} \left( e^{-2eV} D_\alpha e^{2eV} \right) = \tfrac{1}{2} \bar{\sigma}_\nu^{\dot \alpha \alpha} \bar{D}_{\dot \alpha} ( D_\alpha V - e [V, D_\alpha V] ).
\end{equation*}
As can be shown straightforwardly, this transforms under $\delta_\Lambda$ to
\begin{equation*}
- \tfrac{i}{4e} \bar{\sigma}_\nu^{\dot \alpha \alpha} \bar{D}_{\dot \alpha} D_\alpha \Lambda + i [\Lambda, Y_\nu].
\end{equation*}
Using~(\ref{eq:D_anticomm}) and~(\ref{eq:sigma_trace}), we exactly recover~(\ref{eq:delta_Y}). In components, we have
\begin{equation}
\label{eq:Y}
 Y_\nu = A_\nu - i \theta \sigma_\nu \bar \lambda + i \lambda \sigma_\nu \bar \theta + \text{ higher orders in } \theta, \bar \theta.
\end{equation}
Thus, the body of $Y_\nu$ is just $A_\nu$, as one would expect. Therefore, instead of $\int \ud^4q \ f^{\mu \nu}(X) F_{\mu \nu}$, we consider the observable
\begin{equation}
\label{eq:SUSY_F_mu_nu_X}
\int \ud^6q \ f^\alpha(X) W_\alpha + \text{ h.c.},
\end{equation}
with $f^\alpha$ given by~(\ref{eq:f_alpha}), i.e., containing a single $\theta$-component. But, as we saw in~(\ref{eq:Y}), $Y_\nu$ also has a $\theta$-component, $-i\theta \sigma_\nu \bar \lambda$. It follows that $f^\alpha(X)$ has a $\theta^2$-component that involves $\bar \lambda$. This, together with the body of $W_\alpha$, also contributes to~(\ref{eq:SUSY_F_mu_nu_X}). As can be seen from~(\ref{eq:W_alpha_Components}), the body of $W_\alpha$ is $- 2 i \lambda_\alpha$. Thus, the observable~(\ref{eq:SUSY_F_mu_nu_X}) also contains a component with a product of $\bar \lambda$ and $\lambda$. We compute it explicitly for $f^\alpha$ given by~(\ref{eq:f_alpha}), to first order in $e$:
\begin{multline*}
(2\pi)^{-2} \int \ud^4k \ ( \tfrac{i}{2} )( \sigma^{\mu \nu})_\beta^{\ \alpha}  \hat{f}_{\mu \nu}(-k) (i e k\Theta)_\lambda \sigma^\lambda_{\gamma \dot \alpha} \\
\times \int \ud^6q \ \theta^\beta e^{ikq} \theta^\gamma (-i) P_1(k \Theta \del) \bar \lambda^{\dot \alpha} (-2i) \lambda_\alpha + \text{ h.c.}
\end{multline*}
This can be brought to the form
\begin{equation}
\label{eq:SUSY_CovCoor}
\frac{e}{2} (2\pi)^{-2} \int \ud^4k \ \hat{f}^{\mu \nu}(-k) (k\Theta)^\lambda  (\bar \sigma_\lambda \sigma_{\mu \nu} )^{\dotalpha \alpha} \int \ud^4q \ e^{ikq} P_1(k \Theta \del) \bar{\lambda}_{\dot \alpha} \lambda_\alpha + \text{ h.c.}
\end{equation}
Each further $\bar \lambda$ coming in through the covariant coordinate would bring in another $\theta$, so there are no terms with more than two photinos. However, the covariant coordinate can provide for arbitrary powers of $A_\mu$.

As an aside, we mention that the supersymmetric version of the covariant coordinates might be of interest for the noncommutative emergent gravity scenario \cite{Steinacker}.

\section{The two-point function}
\label{sec:2pt}

As discussed in Section~\ref{sec:DispRel}, the dispersion relations can be obtained from the two-point function of the interacting field. The goal of the following sections is thus to compute
\begin{equation}
\label{eq:NCQED2pt}
 \bra{\Omega} \left( \int \ud^4q \ f^{\mu \nu}(X) F_{\mu \nu} \right) \left( \int \ud^4q \ h^{\lambda \rho}(X) F_{\lambda \rho} \right) \ket{\Omega}
\end{equation}
to second order in $e$. In the supersymmetric case the two observables are replaced by the corresponding observables of the form \eqref{eq:SUSY_F_mu_nu_X} with $f^\alpha$ and $h^\alpha$ given by \eqref{eq:f_alpha}. Here we may assume $f^{\mu \nu}$ and $h^{\lambda \rho}$ to be anti-symmetric. Because of the presence of the commutator term of the field strength and the covariant coordinates, a single observable (\ref{eq:NCQEDObservable})
contains, at order $e^n$, $n+1$ photon fields (here we count photinos as photons). Thus, the two-point function (\ref{eq:NCQED2pt}) contains, at order $e^2$, also three- and four-point functions of the photon field. In order to manage the combinatorics, We split the computation of (\ref{eq:NCQED2pt}) into three parts:
\begin{enumerate}

\item We expand the single observable (\ref{eq:NCQEDObservable}) (or \eqref{eq:SUSY_F_mu_nu_X} in the supersymmetric case) in powers of $e$, which is equivalent to an expansion in the number of photon fields. The result is written in the form
\begin{equation*}
 \int \ud^4q \ f^{\mu \nu}(X) F_{\mu \nu} = \int \ud^4k \ \hat f^{\mu \nu}(-k) K_{\mu \nu}(k)
\end{equation*}
where
\begin{equation*}
 K_{\mu \nu}(k) = \sum_{n=1}^\infty e^{n-1} \int \ud^{4n} \underline{k} \ K_{\mu \nu}^{\underline{a}}(k ; \underline{k}) \hat F_{a_1}(k_1) \dots \hat F_{a_n}(k_n).
\end{equation*}
The index $a_i$ stands either for $\mu_i$, $\alpha_i$ or $\dotalpha_i$. Correspondingly $F_{a_i}$ stands for $A_{\mu_i}$, $\lambda_{\alpha_i}$ or $\bar{\lambda}_{\dotalpha_i}$. We used the abbreviations $\underline k = (k_1, \dots k_n)$, $\underline a = (a_1, \dots a_n)$. For our computation, we need the {\it kernels} $K^{\underline{a}}_{\mu \nu}(k ; \underline{k})$ up to $n=3$.

\item We compute the $n$--point functions
\begin{equation*}
 W_{\underline{a}}(\underline{k}) = \bra{\Omega} \hat F_{a_1}(k_1) \dots \hat F_{a_n}(k_n) \ket{\Omega}
\end{equation*}
of the photon field. Here we used the same notation as above. In the following, we call these the \emph{elementary} $n$-point functions. For our computation, we need the elementary two-, three-, and four-point functions to second, first, and zeroth order, respectively.

\item The two-point function (\ref{eq:NCQED2pt}) is now given by
\begin{multline}
\label{eq:2ptCombinatorics}
\sum_{\substack{m=1 \\ n=1}}^\infty e^{m+n-2} \int \ud^4k \ud^4p \ \hat f^{\mu \nu}(-k) \hat h^{\lambda \rho}(-p)  \\ \times \int \ud^{4m} \underline{k} \ud^{4n} \underline{p} \ K_{\mu \nu}^{\underline{a}}(k;\underline{k}) K_{\lambda \rho}^{\underline{b}}(p;\underline{p}) W_{\underline{a} \underline{b}}(\underline{k}, \underline{p}).
\end{multline}
This will be called the \emph{full} two-point function in the following.
\end{enumerate}

This is the program for the next three sections. We remark that (\ref{eq:2ptCombinatorics}) can be straightforwardly generalized for the computation of higher $n$-point functions.

\section{The kernels}

The zeroth order component of $K_{\mu \nu}(k)$ can be directly read off:
\begin{equation}
\label{eq:K0}
 K_{\mu \nu}^{\mu_1}(k;k_1) = - 2 i \delta(k-k_1) k_\mu \delta_\nu^{\mu_1}.
\end{equation}
Here we used the antisymmetry of $f^{\mu \nu}$. At first order, there are two contributions, one from the commutator term in the field strength and one from the covariant coordinate. We have
\begin{multline*}
- i e \int \ud^4q \ f^{\mu \nu}(q) [A_\mu, A_\nu] = - e \frac{1}{(2 \pi)^{2}} \int \ud^4k \ \hat f^{\mu \nu}(-k) \\ \times \int \ud^4k_1 \ud^4k_2 \ \{ \hat A_\mu(k_1), \hat A_\nu(k_2) \} \sin \tfrac{k_1 \Theta k_2}{2} \delta(k-k_1-k_2).
\end{multline*}
In order to find the first order contribution from the covariant coordinate, we have to compute, cf.~(\ref{eq:e_ikX}),
\begin{multline*}
2 \int \ud^4 q \ f^{\mu \nu}(X)|_e \del_\mu A_\nu = e \frac{1}{(2 \pi)^{6}} \int \ud^4k \ud^4k_1 \ud^4k_2 \ \hat f^{\mu \nu}(-k) (i k \Theta)^\rho (-i k_2)_\mu \\ \times \{ \hat A_\rho(k_1), \hat A_\nu(k_2) \} P_1(-ik \Theta k_1) \int \ud^4q \ e^{ikq} e^{-ik_1q} e^{-ik_2q}.
\end{multline*}
The integral over $q$ yields $(2\pi)^4 e^{\frac{i}{2} k \Theta k_1} \delta(k-k_1-k_2)$. Using
\begin{equation*}
 P_1(ix) e^{-\frac{i}{2}x} = \tfrac{\sin x/2}{x/2},
\end{equation*}
we thus find the kernel
\begin{align}
 \label{eq:K1_1}
 K_{\mu \nu}^{\mu_1 \mu_2}(k; k_1, k_2) & = \tfrac{1}{(2\pi)^{2}} \delta(k- \sum k_i) \bigg[ - 2 \delta^{\mu_1}_\mu \delta^{\mu_2}_\nu \sin \tfrac{k_1 \Theta k_2}{2}  \\
\label{eq:K1_2}
 & \qquad +  \left\{ (k \Theta)^{\mu_1} k_{2 \mu} \delta^{\mu_2}_\nu + (k \Theta)^{\mu_2} k_{1 \mu} \delta^{\mu_1}_\nu \right\} \tfrac{\sin \frac{k_1 \Theta k_2}{2}}{\frac{k_1 \Theta k_2}{2}} \bigg]
\end{align}

At second order, there are again two terms. In the same manner as above, we find
\begin{align}
 & K_{\mu \nu}^{\underline \mu}(k; \underline k) = \tfrac{- i}{(2\pi)^{4}} \delta(k- \sum k_i) \nonumber \\
\label{eq:K2_1}
& \times \bigg[ \left\{ (k \Theta)^{\mu_1} \delta^{\mu_2}_\mu \delta^{\mu_3}_\nu \tfrac{\sin \frac{k \Theta k_1}{2}}{\frac{k \Theta k_1}{2}} \sin \tfrac{k_2 \Theta k_3}{2} + (k \Theta)^{\mu_3} \delta^{\mu_1}_\mu \delta^{\mu_2}_\nu \tfrac{\sin \frac{k \Theta k_3}{2}}{\frac{k \Theta k_3}{2}} \sin \tfrac{k_1 \Theta k_2}{2} \right\} \\
\label{eq:K2_2}
 & \quad + \left\{ (k \Theta)^{\mu_1} (k \Theta)^{\mu_2} k_{3 \mu} \delta^{\mu_3}_\nu P_2(- i k \Theta k_1, - i k \Theta k_2) e^{-\frac{i}{2} k_1 \Theta k_2} e^{- \frac{i}{2} (k_1+k_2) \Theta k_3} \right. \\
& \quad \quad \left. + (k \Theta)^{\mu_2} (k \Theta)^{\mu_3} k_{1 \mu} \delta^{\mu_1}_\nu P_2(- i k \Theta k_2, - i k \Theta k_3) e^{-\frac{i}{2} k_2 \Theta k_3} e^{- \frac{i}{2} (k_2+k_3) \Theta k_1} \right\} \bigg]. \nonumber
\end{align}
Here we used again the notation $\underline k = (k_1, k_2, k_3)$. In the first term, one power of $e$ stems from the commutator term of the field strength and the other one from the covariant coordinate. In the second one, the covariant coordinate contributes both powers of $e$.

\subsection{The supersymmetric case}
Because of (\ref{eq:3sigmas_2}), we have
\begin{equation}
\label{eq:2sigmas}
 \bar \sigma_\lambda \sigma_{\mu \nu} = \tfrac{1}{2} \left( - g_{\nu \lambda} \bar \sigma_\mu + g_{\mu \lambda} \bar \sigma_\nu - i \epsilon_{\mu \nu \lambda \kappa} \bar \sigma^\kappa \right).
\end{equation}
When we add the hermitean conjugate in (\ref{eq:SUSY_CovCoor}), the first two terms in (\ref{eq:2sigmas}) drop out (this is due to the presence of $(k \sigma)^\lambda$, which changes sign under conjugation). Employing the symmetrized product of $f^\alpha(X)$ and $W_\alpha$, we obtain the supplementary first order kernels
\begin{subequations}
\label{eq:K1_SUSY}
\begin{align}
\label{eq:K1_3}
 K_{\mu \nu}^{\dotalpha \alpha}(k;k_1,k_2) = & - \tfrac{i}{4} (2\pi)^{-2} \delta(k-k_1-k_2) (k\Theta)^\lambda \epsilon_{\mu \nu \lambda \kappa} ( \bar \sigma^\kappa )^{\dotalpha \alpha} \tfrac{\sin \frac{k \Theta k_1}{2}}{\frac{k \Theta k_1}{2}}, \\
\label{eq:K1_4}
 K_{\mu \nu}^{\alpha \dotalpha}(k;k_1,k_2) = & \tfrac{i}{4} (2\pi)^{-2} \delta(k-k_1-k_2) (k\Theta)^\lambda \epsilon_{\mu \nu \lambda \kappa} ( \bar \sigma^\kappa )^{\dotalpha \alpha} \tfrac{\sin \frac{k \Theta k_1}{2}}{\frac{k \Theta k_1}{2}}.
\end{align}
\end{subequations}
We also find the second order kernels
\begin{subequations}
\begin{align}
\label{eq:K2_3}
 K_{\mu \nu}^{\rho \dotalpha \alpha}(k; \underline k) & = \tfrac{1}{4} (2\pi)^{-4} \delta(k-\sum k_i) (k \Theta)^\rho (k \Theta)^\lambda \epsilon_{\mu \nu \lambda \kappa} ( \bar \sigma^\kappa )^{\dotalpha \alpha} e^{-\frac{i}{2} (k_1+k_2) \Theta k_3} \\
 \times & \left( P_2(-ik\Theta k_1, - i k\Theta k_2) e^{- \frac{i}{2} k_1 \Theta k_2} + P_2(-ik\Theta k_2, - i k\Theta k_1) e^{ \frac{i}{2} k_1 \Theta k_2} \right), \nonumber \\
\label{eq:K2_4}
 K_{\mu \nu}^{\rho \alpha \dotalpha}(k; \underline k) & =  - \tfrac{1}{4} (2\pi)^{-4} \delta(k-\sum k_i) (k \Theta)^\rho (k \Theta)^\lambda \epsilon_{\mu \nu \lambda \kappa} ( \bar \sigma^\kappa )^{\dotalpha \alpha} e^{-\frac{i}{2} (k_1+k_3) \Theta k_2} \\
 \times & \left( P_2(-ik\Theta k_1, - i k\Theta k_3) e^{- \frac{i}{2} k_1 \Theta k_3} + P_2(-ik\Theta k_3, - i k\Theta k_1) e^{ \frac{i}{2} k_1 \Theta k_3} \right). \nonumber
\end{align}
\end{subequations}

\section{The elementary $n$-point functions}
\label{npt}

The next step of the program outlined in Section~\ref{sec:2pt} is the computation of the relevant elementary $n$-point functions. We start with the computation of the elementary two-point function.
At zeroth order in $e$, we have the usual contribution to the photon two-point function:
\begin{equation*}
 W_{\mu \nu}(k,p) = - (2 \pi)^2 g_{\mu \nu} \hat \Delta_+(k) \delta(k+p).
\end{equation*}
There is no first order contribution. At second order, there are three terms,
\begin{equation}
\label{eq:NCQEDThreeTerms}
 \bra{\Omega} A_{2 \mu}(k) A_{0 \nu}(p) \ket{\Omega} + \bra{\Omega} A_{0 \mu}(k) A_{2 \nu}(p) \ket{\Omega} + \bra{\Omega} A_{1 \mu}(k) A_{1 \nu}(p) \ket{\Omega},
\end{equation}
similarly to the case of the $\phi^3$ model, cf. \cite{NCDispRel}.
As discussed in Section~\ref{sec:DispRel}, we treat the sum of the first two terms by computing the self-energy\footnote{We determine the self-energy by writing the contracted part of $\hat{A}_2^\mu(k)$ as $(2\pi)^2 \Pi^{\mu \nu}(k) \hat{A}_{0 \nu}(k)$.} $\Pi_{\mu \nu}(k)$ and then setting
\begin{equation}
\label{eq:W_SelfEnergy}
 W_{\mu \nu}(k,p) = - (2\pi)^2 \delta(k+p) \Pi_{\mu \nu}(k) \tfrac{\del}{\del m^2} \hat \Delta_+(k).
\end{equation}
As shown in \cite{NCDispRel}, this follows from a properly defined adiabatic limit if $\Pi_{\mu \nu}(-k) = \Pi_{\mu \nu}(k)$. In the present (massless) case, this condition is too strong as we will see below. However, as shown in \cite[Remark 5.2.4]{diss}, $\Pi_{\mu \nu}(k) - \Pi_{\mu \nu}(-k) \propto k^4$ is sufficient for the adiabatic limit to be well-defined and (\ref{eq:W_SelfEnergy}) to be applicable.

The third term in (\ref{eq:NCQEDThreeTerms}) is represented by the two graphs
\begin{center}
\includegraphics{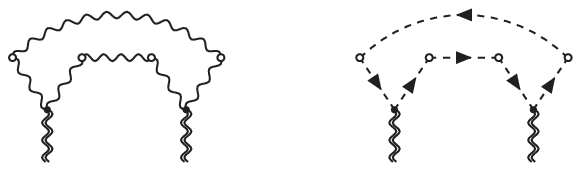}
\end{center}
and contributes to the continuous spectrum. It is thus not relevant for our discussion, since we are mainly interested in the distortion of the dispersion relations. For the computation of the above graphs, we thus refer to \cite[Section 6.8.1]{diss}.

The distortion of the dispersion relation is due to the self-energy, i.e., the first two terms in (\ref{eq:NCQEDThreeTerms}). We start with the computation of the tadpole, i.e., the following graphs:
\begin{center}
\includegraphics{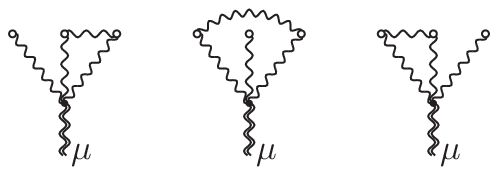}
\end{center}
For the first and the third graph, we find
\begin{equation*}
- e^2 (d-1) \hat A^\mu_0 (k) \hat \Delta_R(k) \int \ud^4 l \ \hat \Delta_+(l) \sin^2 \tfrac{k \Theta l}{2},
\end{equation*}
respectively. For the second graph, we find this twice. Thus, the tadpole contribution to the self-energy is
\begin{equation*}
 \Pi_{\mu \nu}(k) = - 4 (2\pi)^{-2} e^2 (d-1) g_{\mu \nu} \int \ud^4 l \ \hat \Delta_+(l) \sin^2 \tfrac{k \Theta l}{2}.
\end{equation*}
In the following, we write all the contributions to the self-energy in the form
\begin{equation}
\label{eq:Pi}
  \Pi_{\mu \nu}(k) = 2 e^2 \int \ud^4l \ \hat{\Delta}^{(1)}(l) \hat{\Delta}_R(k-l) \sin^2 \tfrac{k \Theta l}{2}  \pi_{\mu \nu}(k,l),
\end{equation}
with
\begin{equation*}
 \Delta^{(1)}(x) = \Delta_+(x) + \Delta_+(-x).
\end{equation*}
For the tadpole this can be done using $k^2 \hat{\Delta}_R(k) = - (2\pi)^{-2}$ and the symmetry of $\hat \Delta^{(1)}$:
\begin{equation*}
 \pi^{\text{tp}}_{\mu \nu}(k,l) = (d-1) g_{\mu \nu} (k-l)^2.
\end{equation*}

Now we come to the photon fish graph. We have to compute the following graphs:
\begin{center}
\includegraphics{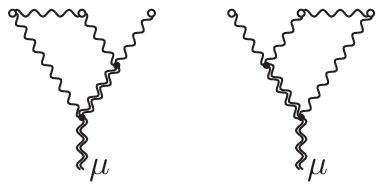}
\end{center}
These have to be counted twice in order to account for the other possible contraction. For the first graph one obtains
\begin{multline*}
 - e^2 (2 \pi)^2 \hat A^\nu_0 (k) \hat \Delta_R(k) \int \ud^4 l \ \hat \Delta_+(l) \hat \Delta_R(k-l) \sin^2 \tfrac{k \Theta l}{2} \\
 \times \left\{ g_{\mu \nu} (5 k^2 - 2 k \cdot l) + (d-6) k_\mu  k_\nu + (2d-3) \left( 2 l_\mu l_\nu - k_\mu l_\nu - l_\mu k_\nu \right) \right\}.
\end{multline*}
For the second graph, one finds nearly the same expression, but with $\hat \Delta_+(l)$ replaced by $\hat \Delta_+(-l)$. Taking the factor $2$ into account, we thus obtain for the photon fish graph
\begin{equation*}
 \pi^{\text{pf}}_{\mu \nu}(k,l) = - g_{\mu \nu} (5 k^2 - 2  k \cdot l ) - (d-6) k_{\mu} k_{\nu} + (2d-3) \left\{ (k-l)_{\mu} l_{\nu} + l_{\mu} (k-l)_{\nu} \right\}.
\end{equation*}

It remains to treat the ghost fish graphs:
\begin{center}
\includegraphics{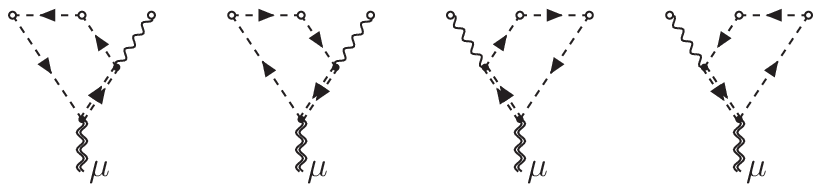}
\end{center}
We have to count these graphs twice, in order to account for the case where the photon leaves the second vertex to the other side.
For the first graph, one obtains
\begin{equation*}
- e^2 (2 \pi)^2 \hat A^\nu_0 (k) \hat \Delta_R(k) \int \ud^4 l \ \hat \Delta_+(l) \hat \Delta_R(k-l) \sin^2 \tfrac{k \Theta l}{2} l_\nu (k-l)_\mu.
\end{equation*}
The second graph yields a similar expression, but with $\mu$ and $\nu$ interchanged in the loop integral. For the last two graphs, one finds the same results, but with $\hat \Delta_+(l)$ replaced by $\hat \Delta_+(-l)$. Thus, the ghost loop contribution is
\begin{equation*}
 \pi^{\text{gh}}_{\mu \nu}(k,l) = - (k-l)_{\mu} l_{\nu} - l_{\mu} (k-l)_{\nu}.
\end{equation*}
Adding all this up, we find, using $l^2 \hat{\Delta}^{(1)}(l) = 0$,
\begin{subequations}
\begin{align}
\label{subeq:f_tot_1}
 \pi^{\text{tot}}_{\mu \nu}(k,l) = & - (6-d) \left( g_{\mu \nu} k^2 - k_{\mu} k_{\nu} \right) \\ 
\label{subeq:f_tot_2}
  & - 2 (d-2) \left( g_{\mu \nu} k \cdot l - (k-l)_{\mu} l_{\nu} - l_{\mu} (k-l)_{\nu} \right).
\end{align}
\end{subequations}
In the remainder of this subsection, we want to compute $\Pi_{\mu \nu}(k)$ explicitly. We do this separately for the planar and the nonplanar part, where the split is given by
\begin{equation}
\label{eq:SinSplit}
 \sin^2 \tfrac{k \Theta l}{2} = \tfrac{1}{2}-\tfrac{1}{2} \cos k \Theta l,
\end{equation}
in the sense that the first term on the \rhs gives rise to the planar part and the second term to the nonplanar part.

\subsection{The planar part}

We first focus on the term~(\ref{subeq:f_tot_1}). It already has the usual tensor structure. The loop integral is the same as that of the fish graph in the massless $\phi^3$ model and corresponds, in position space, to computing the point-wise product $\Delta^{(1)} \Delta_R$.
Using \cite{BDFP02}
\begin{equation*}
 \Delta^{(1)} \Delta_{R} = - i \Delta_F \Delta_F - i \Delta_{-} \Delta_{-},
\end{equation*}
where $\Delta_-(x) = \Delta_+(-x)$, and the well-known expression for the square of the Feynman propagator (see, e.g., \cite{Itzy}), we obtain, after renormalization, in the limit $k^2 \to 0$,
\begin{equation}
\label{eq:planarIR1}
 \Pi^{\text{a,pl}}_{\mu \nu}(k) = e^2 (2\pi)^{-2} ( g_{\mu \nu} k^2 - k_\mu k_\nu ) \left( \ln \tfrac{\sqrt{k^2}}{\mu} - \theta(k^2) \varepsilon(k_0) \tfrac{i \pi}{2} \right).
\end{equation}
Here $\mu$ is some mass scale that depends on the renormalization condition, and $\varepsilon$ is the sign function.

In order to treat the term~(\ref{subeq:f_tot_2}) in a similar fashion, we need the identity
\begin{equation*}
 \del_\mu \Delta_R(x) = \del_\mu \left( \theta(x^0) \Delta(x) \right) = \theta(x^0) \del_{\mu} \Delta(x).
\end{equation*}
Here we used $\delta(x^0) \Delta(x) = 0$. Similarly, we obtain
\begin{equation*}
 \del_\mu \Delta_F(x) = \theta(x^0) \del_\mu \Delta_+(x) + \theta(-x^0) \del_\mu \Delta_-(x).
\end{equation*}
One can now show that
\begin{multline*}
 g^{\mu \nu} \del_\lambda \Delta_R \del^\lambda \Delta^{(1)} - \del^\mu \Delta_R \del^\nu \Delta^{(1)} - \del^\nu \Delta_R \del^\mu \Delta^{(1)} \\ = - i g^{\mu \nu} \del_\lambda \Delta_F \del^\lambda \Delta_F + 2 i \del^\mu \Delta_F \del^\nu \Delta_F - i g^{\mu \nu} \del_\lambda \Delta_- \del^\lambda \Delta_- + 2 i \del^\mu \Delta_- \del^\nu \Delta_-
\end{multline*}
holds. 
The first two terms on the \rhs are terms that one also finds in the two-point function of a nonabelian gauge theory. As there, their sum can be renormalized in such a way that the Ward identity (transversality) is fulfilled. One then obtains, in the limit $k^2 \to 0$, the contribution
\begin{equation}
\label{eq:planarIR2}
 \Pi^{\text{b,pl}}_{\mu \nu}(k) = \tfrac{2}{3} (2\pi)^{-2} e^2 ( g_{\mu \nu} k^2 - k_\mu k_\nu ) \left( \ln \tfrac{\sqrt{k^2}}{\mu} - \theta(k^2) \varepsilon(k_0) \tfrac{i \pi}{2} \right).
\end{equation}
There are some potential infrared problems arising from (\ref{eq:planarIR1}) and (\ref{eq:planarIR2}): The expression is not well-defined for $k^2 \to 0$. Furthermore, $\Pi_{\mu \nu}(k)$ and $\Pi_{\mu \nu}(-k)$ do not coincide in a neighborhood of the forward light cone, because of the imaginary part. Then, as discussed below (\ref{eq:W_SelfEnergy}), the adiabatic limit is not well-defined and (\ref{eq:W_SelfEnergy}) is not applicable. Such difficulties are typical for nonabelian gauge theories. We come back to this problem later.

\subsection{The nonplanar part}
\label{sec:nonplanar}

Now we take a look at the nonplanar part. The loop integral corresponding to the term~(\ref{subeq:f_tot_1}) is given by
\begin{equation*}
 \Pi^{\text{a,np}}_{\mu \nu}(k) = 2 e^2 (g_{\mu \nu} k^2 - k_\mu k_\nu)  \int \ud^4l \ \hat \Delta_+(l) \cos k \Theta l \left( \hat \Delta_R(k-l) + \hat \Delta_R(k+l) \right),
\end{equation*}
which is the same as for the massless $\phi^3$ fish graph. For the massive case, this integral was defined and computed in the sense of oscillatory integrals in \cite{NCDispRel}. Unfortunately, this does not seem to be possible in the massless case. Thus, we follow a different route and interpret $y = k\Theta$ as an independent variable. Then the loop integral is a distribution $F(y,k)$ in two variables,
\begin{equation*}
 F(y,k) = 2 \int \ud^4l \ \hat \Delta_+(l) \cos yl \left( \hat \Delta_R(k-l) + \hat \Delta_R(k+l) \right).
\end{equation*}
This is well-defined as a tempered distribution in the two variables. Furthermore, as shown in \cite{diss}, it can be restricted to $y = k \Theta$ for $k^2>0$, so that the loop integral exists in this sense. Here, we compute it formally, obtaining the same result.
We are only interested in the behaviour for $k^2>0$. Due to Lorentz invariance, we may choose $k=(k_0, \V{0})$. Then we have
\begin{align*}
 F(y,k) & = \frac{-2}{(2 \pi)^{3}} \int \frac{\ud^3 \V{l}}{2 l} \ \left( \frac{\cos (y_0 l - \V{y} \cdot \V{l})}{k^2 - 2 k_0 l + i \epsilon (k_0-l)} + \frac{\cos (y_0 l - \V{y} \cdot \V{l})}{k^2 + 2 k_0 l + i \epsilon (k_0+l)} \right) \\
 & = \frac{-4}{(2 \pi)^{3}} \int \frac{\ud^3 \V{l}}{2 l} \ \frac{1}{k^2 - 4 l^2 + i \epsilon k_0} \cos (y_0 l - \V{y} \cdot \V{l}) \\
 & = \frac{-2}{\betrag{\V{y}} (2 \pi)^{2}} \int_0^{\infty} \ud l \ \frac{1}{k^2 - 4 l^2 + i \epsilon k_0}  \left\{ \sin [ ( \betrag{\V y} + y_0 ) l ] + \sin [ ( \betrag{\V y} - y_0 ) l ] \right\}.
\end{align*}
We define
\begin{align*}
 G_0^\pm(a,b,c) & = \frac{1}{a} \int_0^\infty \ud l \ \frac{\sin [(a+b)l]}{c^2 \pm i \epsilon - l^2} \\
 & = \mp \frac{i \pi}{2} \frac{\sin (a+b)c}{ac} + \frac{1}{a} \int_0^\infty \ud l \ \sin (a+b)l \ \pv{c^2 - l^2}.
\end{align*}
The second term is a standard integral. With \cite[Eq.~(3.723.8)]{Gradshteyn}, we obtain
\begin{multline*}
 G_0^\pm(a,b,c) = \mp \frac{i \pi}{2} \frac{\sin (a+b)c}{ac} \\ + \frac{1}{ac} \left[ \sin ((a+b)c) \ci ((a+b)c) - \cos ((a+b)c) \si ((a+b)c)  \right].
\end{multline*}
Here $\si$ and $\ci$ are the sine and cosine integral, cf. \cite[8.230]{Gradshteyn}. We have
\begin{equation*}
 F(y,k)= - \tfrac{1}{2} (2\pi)^{-2} \left\{ G_0^{\varepsilon(k_0)}(\betrag{\V y}, y_0, \tfrac{1}{2} \sqrt{k^2}) + G_0^{\varepsilon(k_0)}(\betrag{\V y}, - y_0, \tfrac{1}{2} \sqrt{k^2}) \right\}.
\end{equation*}
where $\varepsilon$ is the sign function.
We write the nonplanar contribution of the term~(\ref{subeq:f_tot_1}) to the self-energy in the form
\begin{equation*}
 \Pi^{\text{a,np}}_{\mu \nu} (k) = \left( g_{\mu \nu} k^2 -k_{\mu} k_{\nu} \right) \Sigma_0(k).
\end{equation*}
and thus obtain, for $k$ timelike,
\begin{equation}
\label{eq:Sigma0}
 \Sigma_0(k) = - (2\pi)^{-2} G_0^{\varepsilon(k_0)}(\sqrt{-(k \Theta)^2}, 0, \tfrac{1}{2} \sqrt{k^2}).
\end{equation}

It remains to treat the terms~(\ref{subeq:f_tot_2}). We define
\begin{equation*}
F_{\mu \nu}(y,k) = 4 \int \ud^4l \ \hat \Delta^{(1)}(l) \hat \Delta_R(k-l) \cos yl \left\{ g_{\mu \nu} k \cdot l - (k-l)_\mu l_\nu - l_\mu (k-l)_\nu \right\}.
\end{equation*}
As shown in \cite{diss}, $F_{\mu \nu}$ can be expressed by derivatives of $F$. The nonplanar contribution of \eqref{subeq:f_tot_2} to the self-energy is given by $\Pi^{\text{b,np}}_{\mu \nu}(k) = e^2 F_{\mu \nu}(k\Theta, k)$. One can show that, for timelike $k$, it is of the form
\begin{equation*}
 \Pi^{\text{b,np}}_{\mu \nu} (k) = e^2 \left( g_{\mu \nu} k^2 -k_{\mu} k_{\nu} \right) \Sigma_1(k) + e^2 \tfrac{(k \Theta)_{\mu} (k \Theta)_{\nu}}{(k \Theta)^4} \Sigma_2(k),
\end{equation*}
with
\begin{align*}
 \Sigma_1(k) & = - \tfrac{4}{(2\pi)^{2} k^2} \left\{ - G_2^{\varepsilon(k_0)}(\sqrt{-(k\Theta)^2}, 0, \tfrac{1}{2} \sqrt{k^2}) \right. \\
  & \qquad \qquad \qquad + \tfrac{1}{\sqrt{-(k\Theta)^2}} G_1^{\varepsilon(k_0)}(\sqrt{-(k\Theta)^2}, 0, \tfrac{1}{2} \sqrt{ k^2}) \\
  & \qquad \qquad \qquad \left. + \tfrac{1}{(k\Theta)^2} G_0^{\varepsilon(k_0)}(\sqrt{-(k\Theta)^2}, 0, \tfrac{1}{2} \sqrt{k^2}) \right\} , \\
 \Sigma_2(k) & = \tfrac{4 (k \Theta)^2}{(2\pi)^2} \left\{ - G_2^{\varepsilon(k_0)}(\sqrt{-(k\Theta)^2}, 0, \tfrac{1}{2} \sqrt{k^2}) \right. \\
  & \qquad \qquad \quad + \tfrac{3}{\sqrt{-(k\Theta)^2}} G_1^{\varepsilon(k_0)}(\sqrt{-(k\Theta)^2}, 0, \tfrac{1}{2} \sqrt{k^2})  \\
  & \qquad \qquad \quad \left. + \tfrac{3}{(k\Theta)^2} G_0^{\varepsilon(k_0)}(\sqrt{-(k\Theta)^2}, 0, \tfrac{1}{2} \sqrt{k^2}) \right\}.
\end{align*}
Here, $G^\pm_n$ is given by 
\begin{equation*}
 G_n^\pm(a,b,c)=\frac{\del^n}{\del b^n} G_0^\pm(a,b,c). 
\end{equation*}
We emphasize that the nonplanar loop integrals are completely (analytically) solved for $k^2>0$. To the best of our knowledge, this has not been achieved before. In the literature, only the leading behaviour for $(k\Theta)^2 \to 0$ was computed. We want to compare with these results. Using the series expansions \cite[8.232]{Gradshteyn} of $\si$ and $\ci$, one finds
\begin{align}
\label{eq:nonplanarIR1}
 \Sigma_0(k) = & \tfrac{-1}{(2\pi)^{2}} \left( \ln \tfrac{1}{2} \sqrt{-(k\Theta)^2} \sqrt{k^2} +\gamma-1 - \varepsilon(k_0) \tfrac{i\pi}{2} \right) + \order((k\Theta)^2 k^2), \\
\label{eq:nonplanarIR2}
 \Sigma_1(k) = & \tfrac{-1}{(2\pi)^{2}} \left( \tfrac{2}{3} \ln \tfrac{1}{2} \sqrt{-(k\Theta)^2} \sqrt{k^2} + \tfrac{2}{3} \gamma - \tfrac{5}{9} - \varepsilon(k_0) \tfrac{2}{3} \tfrac{i\pi}{2} \right) + \order((k\Theta)^2 k^2), \\
\label{eq:nonplanarIR3}
 \Sigma_2(k) = & \tfrac{-1}{(2\pi)^{2}} \left( 8 - \tfrac{1}{3} (k\Theta)^2 k^2 \right) + \order((k\Theta)^4 k^4).
\end{align}
This is in agreement with the results obtained in \cite{hayakawa} in the setting of the modified Feynman rules, cf.~(\ref{eq:Hayakawa_Sigma_1},b). Adding~(\ref{eq:nonplanarIR1}) and~(\ref{eq:nonplanarIR2}) to the planar terms (\ref{eq:planarIR1}) and (\ref{eq:planarIR2}), we obtain the following contribution to the self-energy
\begin{equation*}
 \Pi_{\mu \nu}(k) = - \tfrac{5}{3} (2\pi)^{-2} e^2 (g_{\mu \nu} k^2 - k_\mu k_\nu) \left( \ln \mu \sqrt{-(k \Theta)^2} + \order( k^2 (k \Theta)^2) \right).
\end{equation*}
We see that the problematic term proportional to $\ln k^2$ drops out. Also the imaginary parts cancel between the planar and the nonplanar part at zeroth order in $k^2$. This cancellation is due to the fact that we compute the difference between the planar and the nonplanar part, cf. (\ref{eq:SinSplit}), which is a consequence of the interaction term being a commutator\footnote{To the best of our knowledge, this cancellation of infrared divergences between the planar and nonplanar parts has not been noticed before.}. As discussed below (\ref{eq:W_SelfEnergy}), we need $\Pi_{\mu \nu}(k) - \Pi_{\mu \nu}(-k) \sim k^4$ in the neighborhood of the light cone in order to have a well-defined adiabatic limit. Since this difference is given by the imaginary part, which is of order $k^2$, the condition is fulfilled for the part proportional to $g_{\mu \nu} k^2$. The term proportional to $k_\mu k_\nu$ drops out at order $e^2$, because of the test functions being antisymmetric and the form \eqref{eq:K0} of the zeroth order kernel (for a detailed discussion, we refer to \cite[Section 6.8.2]{diss}.

As is obvious from (\ref{eq:nonplanarIR3}), the imaginary part of $\Sigma_2(k)$ is proportional to $k^4$. Thus, the condition $\Pi_{\mu \nu}(k) - \Pi_{\mu \nu}(-k) \sim k^4$ is also fulfilled for this term and we obtain the following second order contribution to the elementary two-point function:
\begin{align}
\label{eq:W1}
 W_{\mu \nu}(k,p) & = - \tfrac{5}{3} e^2 g_{\mu \nu}  \ln \mu \sqrt{-(k \Theta)^2}  \hat \Delta_+(k) \delta(k+p) \\
\label{eq:W2}
  & \quad - e^2 \delta(k+p) \tfrac{(k\Theta)_\mu (k\Theta)_\nu}{(k\Theta)^4} \left( 8 \tfrac{\del}{\del m^2} \hat \Delta_+(k) - \tfrac{(k \Theta)^2}{3} \hat \Delta_+(k) \right).
\end{align}
Here we used $k^2 \frac{\del}{\del m^2} \hat \Delta_+(k) = \hat \Delta_+(k)$.
The well-definedness of the products of distributions in \eqref{eq:W1} and\eqref{eq:W2} will be discussed later in Section~\ref{sec:infrared}.

\begin{remark}
\label{rem:SigmaSpacelike}
Above, we computed the self-energy $\Pi_{\mu \nu}(k)$ on the light cone by computing it in the interior and then taking the limit $k^2 \to 0$. In principle one should check whether one gets the same result by continuation from $k^2<0$. The knowledge of $\Pi_{\mu \nu}(k)$ for spacelike $k$ is also important if one wants to consider higher loop orders. However, it was not yet possible to rigorously compute the nonplanar part of $\Pi_{\mu \nu}(k)$ in this range. In \cite[Appendix B.8]{diss}, it is shown that a formal calculation of $\Sigma_0(k)$ for $k^2 < 0$, $(k \Theta)^2 > 0$ leads to
\begin{equation}
\label{eq:SigmaSpacelike}
 \Sigma_0(k) = - (2\pi)^{-2} \Re G_0^+(\sqrt{(k \Theta)^2},0, \tfrac{1}{2} \sqrt{-k^2}).
\end{equation}
This is the real part of the expression for timelike $k$, cf. (\ref{eq:Sigma0}), which may be seen as an indication that the approach taken here is consistent.
\end{remark}

\subsection{The elementary three- and four-point functions}

We need the elementary three-point function at first order. Using the cubic photon vertex (\ref{eq:CubicVertex}), it is straightforward to obtain
\begin{align}
\label{eq:3pt}
 W_{\underline \mu}(\underline k) & =  2 i e (2\pi)^4  \sin \tfrac{k_2 \Theta k_3}{2} \delta(\sum k_i)
 \\ & \times \left( g_{\mu_1 \mu_2} (k_1-k_2)_{\mu_3} + g_{\mu_3 \mu_1} (k_3-k_1)_{\mu_2} + g_{\mu_2 \mu_3} (k_2-k_3)_{\mu_1} \right) \nonumber 
 \\ & \times \left( \hat \Delta_R(k_1) \hat \Delta_+(-k_2) \hat \Delta_+(-k_3) + \hat \Delta_+(k_1) \hat \Delta_R(k_2) \hat \Delta_+(-k_3) \right. \nonumber \\
 & \quad \left.  +\hat \Delta_+(k_1) \hat \Delta_+(k_2) \hat \Delta_R(k_3) \right). \nonumber
\end{align}
The elementary four-point function is only needed at zeroth order: 
\begin{align}
\label{eq:4pt}
 W_{\underline \mu}(\underline k) = (2 \pi)^4 & \left( g_{\mu_1 \mu_2} g_{\mu_3 \mu_4} \delta(k_1+k_2) \delta(k_3+k_4) \hat \Delta_+(k_1) \hat \Delta_+(k_3) \right. \\
 & + g_{\mu_1 \mu_3} g_{\mu_2 \mu_4} \delta(k_1+k_3) \delta(k_2+k_4) \hat \Delta_+(k_1) \hat \Delta_+(k_2) \nonumber \\
 & \left. + g_{\mu_1 \mu_4} g_{\mu_2 \mu_3} \delta(k_1+k_4) \delta(k_2+k_3) \hat \Delta_+(k_1) \hat \Delta_+(k_2) \right). \nonumber
\end{align}

\subsection{The supersymmetric case}
As above, we skip the computation of the graph
\begin{center}
\includegraphics{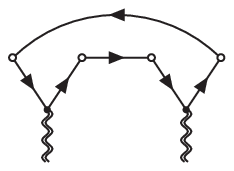}
\end{center}
which only contributes to the two-particle spectrum (the interested reader is referred to \cite{diss}). The contributions that we are interested in come from the self-energy, i.e., the graphs
\begin{center}
\includegraphics{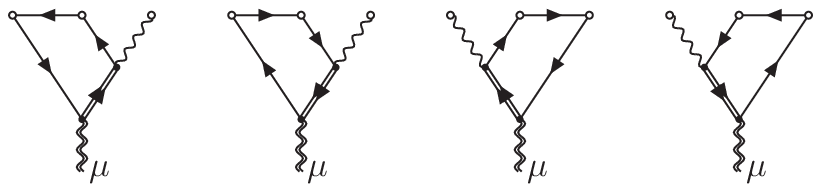}
\end{center}
These have to be counted twice in order to account for the graphs where the photon leaves the second vertex to the other side. With this factor, we obtain, in total,
the following contribution to the self-energy:
\begin{equation*}
 \pi^{\text{ino}}_{\mu \nu}(k,l) = - (k-l)^\lambda l^\rho \left\{ \tr ( \bar \sigma_\lambda \sigma_\mu \bar \sigma_\rho \sigma_\nu ) + \tr ( \bar \sigma_\rho \sigma_\mu \bar \sigma_\lambda \sigma_\nu ) \right\}.
\end{equation*}
Because of~(\ref{eq:sigma_trace}) and~(\ref{eq:3sigmas_1}), we have
\begin{equation}
\label{eq:sigmaTrace}
 \tr ( \sigma^\kappa \bar \sigma^\lambda \sigma^\mu \bar \sigma^\nu ) = 2 \left( - g^{\kappa \mu} g^{\lambda \nu} +  g^{\lambda \mu} g^{\kappa \nu} + g^{\kappa \lambda} g^{\mu \nu} + i \epsilon^{\kappa \lambda \mu \nu} \right).
\end{equation}
In the sum above the imaginary parts cancel each other, and we obtain
\begin{equation*}
 \pi^{\text{ino}}_{\mu \nu}(k,l) = 4 g_{\mu \nu} k \cdot l - 4 (k-l)_\mu l_\nu - 4 (k-l)_\mu l_\nu.
\end{equation*}
This cancels the term (\ref{subeq:f_tot_2}).
Therefore we have shown that the term $\Sigma_2$ vanishes upon introducing supersymmetry, as in the setting of the modified Feynman rules \cite{Matusis}. The final expression for the self-energy in the supersymmetric case is thus
\begin{equation}
\label{eq:Pi_final}
 \Pi_{\mu \nu}(k) = - (2\pi)^{-2} e^2 (g_{\mu \nu} k^2 - k_\mu k_\nu) \left( \ln \mu \sqrt{-(k \Theta)^2} + \order( k^2 (k \Theta)^2) \right).
\end{equation}

The supersymmetric contribution to the elementary three-point function is given by
\begin{align}
\label{eq:W3_SUSY_1}
 W_{\dotalpha \alpha \mu}(\underline k) & = 2 i e (2\pi)^4 (\sigma_\rho \bar \sigma_\mu \sigma_\lambda)_{\alpha \dotalpha} k_1^\lambda k_2^\rho \sin \tfrac{k_2 \Theta k_3}{2} \delta(\sum k_i) \\
& \times \left( \left[ \hat \Delta_R(k_1) \hat \Delta_+(-k_2) + \hat \Delta_+(k_1) \hat \Delta_R(k_2) \right] \right. \nonumber \\
& \quad \left. + \hat \Delta_+(-k_3)  \hat \Delta_+(k_1) \hat \Delta_+(k_2) \hat \Delta_R(k_3) \right). \nonumber
\end{align}
The expressions for the other orders of the fields are obtained by analogously permuting the $k_i$'s in the first line, with an additional sign change for a commutation of $\alpha$ and $\dotalpha$.
We also have the elementary four-point functions
\begin{align}
\label{eq:W4_SUSY_1}
 W_{\dotalpha \alpha \mu \nu}(\underline k) & = - (2 \pi)^4 g_{\mu \nu} \sigma^\lambda_{\alpha \dotalpha} k_{1 \lambda} \hat \Delta_+(k_1) \hat \Delta_+(k_3) \delta(k_1+k_2) \delta(k_3+k_4), \\
 W_{\dotalpha \alpha \dotbeta \beta}(\underline k) & = (2 \pi)^4 \left( \sigma^\lambda_{\alpha \dotalpha} k_{1 \lambda} \sigma^\rho_{\beta \dotbeta} k_{3 \rho} \hat \Delta_+(k_1) \hat \Delta_+(k_3) \delta(k_1+k_2) \delta(k_3+k_4) \right. \nonumber \\
\label{eq:W4_SUSY_2}
 & \left. + \sigma^\lambda_{\alpha \dotbeta} k_{1 \lambda} \sigma^\rho_{\beta \dotalpha} k_{2 \rho} \hat \Delta_+(k_1) \hat \Delta_+(k_2) \delta(k_1+k_4) \delta(k_2+k_3) \right). 
\end{align}

\section{The full two-point function}
\label{sec:Full2pt}

Having calculated $K_{\mu \nu}(k)$ and the relevant elementary $n$-point functions, we now come to the third point of our list in Section~\ref{sec:2pt}, the computation of the full two-point function (\ref{eq:NCQED2pt}). At zeroth order, we find
\begin{equation*}
 - 4 (2\pi)^2 \int \ud^4 k \ \hat f^{\mu \nu}(-k) \hat h^\lambda_{\ \nu}(k) k_\mu k_\lambda \hat \Delta_+(k). 
\end{equation*}
There is no first order contribution. In the following, we compute all second order terms. Since they are not the main point of interest, the loop integrals leading to finite contributions to the continuous spectrum will not be computed explicitly.
We order the presentation of the various contributions by the powers of $e$ that the elementary two-point function $W_{\underline \mu}(\underline k)$, i.e., the interaction, contributes. In order to simplify the notation, we will write the different contributions in the form
\[
e^2 \int \ud^4k \ \hat{f}^{\mu \nu}(-k) \hat{h}^{\lambda \rho}(k) \Gamma_{\mu \nu \lambda \rho}(k)
\]
and only give the corresponding $\Gamma$. The calculations are straightforward, but tedious, so we only present the results and refer the interested reader to \cite[Section 6.9]{diss}.

\subsection{$e^0$ from the interaction}
\label{sec:FreePart}

If the elementary $n$-point function does not contribute any power of $e$, both powers of $e$ must stem from the two kernels $K_{\mu \nu}^{\underline \mu}(k;\underline k)$.
In the case where both kernels are of first order, i.e., $m=n=2$ in (\ref{eq:2ptCombinatorics}), one only finds contributions to the continuous spectrum, which are not relevant for the present discussion.

The terms in which the term (\ref{eq:K2_1}) of the second order kernel and the zeroth order kernel (\ref{eq:K0}) are used vanish.


Finally, we consider the terms involving the term (\ref{eq:K2_2}) of the second order kernel. Using it in the first observable, we find 
\begin{multline*}
 \Gamma_{\mu \nu \lambda \rho}(k) = - 2 k_\lambda \hat \Delta_+(k) \int \ud^4l \ \hat \Delta_+(l) \Big\{ 2 g_{\nu \rho} k_\mu (k \Theta)^2 P_2(-ik \Theta l, ik \Theta l) \\
  + (k \Theta)_\nu (k \Theta)_\rho l_\mu \left[ P_2(ik \Theta l, 0) e^{-ik \Theta l} + P_2(0, ik \Theta l) - (l \leftrightarrow -l) \right] \Big\}.
\end{multline*}
Using
\begin{equation*}
 P_2(x,0) = \frac{1-e^x+x e^x}{x^2}, \quad P_2(0,x) = \frac{-1+e^x-x}{x^2},
\end{equation*}
one can show that the expression in square brackets vanishes. Thus, we are left with
\begin{equation}
\label{eq:CovCoorO2_1_result}
\Gamma_{\mu \nu \lambda \rho}(k) = - 4 g_{\nu \rho} (k \Theta)^2  k_{\mu} k_{\lambda}  \hat{\Delta}_+(k) \int \ud^4 l \ \hat{\Delta}_+(l) P_2(0, -i k \Theta l).
\end{equation}
Here we used $P_2(x,-x) = P_2(0,x)$.
From the term where (\ref{eq:K2_2}) is used in the observable involving $h^{\lambda \rho}$, one obtains a similar contribution, but with $P_2(0,-ik \Theta l)$ replaced by $P_2(0,ik \Theta l)$. We have
\begin{equation*}
 P_2(0,i x) + P_2(0,-ix) = 2 \frac{1 - \cos x}{x^2} = \left( \frac{\sin \frac{x}{2}}{\frac x 2} \right)^2.
\end{equation*}
The combination of both terms thus gives
\begin{equation}
\label{eq:divergence1}
\Gamma_{\mu \nu \lambda \rho}(k) = - 4 g_{\nu \rho} (k \Theta)^2  k_{\mu} k_{\lambda}  \hat{\Delta}_+(k) \int \ud^4 l \ \hat{\Delta}_+(l) \frac{\sin^2 \frac{k \Theta l}{2}}{(\frac{k \Theta l}{2})^2}.
\end{equation}
The integral over $l$ can not be split into a planar (local) and a finite nonplanar part. There is no obvious way to define it rigorously. We want to compute it at least formally. We choose $k=(k_0,\V 0)$ and obtain
\begin{multline*}
 2\pi \int_0^\infty \ud l \ \frac{l}{2} \int_{-1}^1 \ud x \ \frac{\sin^2 \frac{\betrag{ \underline{k \Theta} } l x}{2}}{\left( \frac{\betrag{ \underline{k \Theta} } l x}{2} \right)^2} \\ = 8\pi \int_0^\infty \ud l \ \frac{l}{2} \frac{-1 + \cos (\betrag{ \underline{k \Theta} } l) + \betrag{ \underline{k \Theta} } l \ \si (\betrag{ \underline{k \Theta} } l)}{(\betrag{ \underline{k \Theta} } l)^2}.
\end{multline*}
Here $\underline{k \Theta}$ is the spatial part of $k \Theta $. Because of the last term in the numerator, this diverges linearly. Introducing a cutoff $\Lambda$, this leading divergence leads to
\begin{equation*}
\Gamma_{\mu \nu \lambda \rho}(k) \sim g_{\nu \rho} \sqrt{(k \Theta)^2}  k_{\mu} k_{\lambda}  \hat{\Delta}_+(k) \Lambda.
\end{equation*}
As the momentum appears in a square root, we are faced with a nonlocal divergence.
We remark that this term is obtained from the contraction of the two fields stemming from the covariant coordinates. Thus, subtracting it can be interpreted as a normal ordering of $e^{i k X}$. This amounts to a redefinition $f^{\mu \nu} \to f^{\mu \nu} + f_c^{\mu \nu}$ with
\begin{equation*}
 \hat{f}^{\mu \nu}_c(-k) = (2 \pi)^{-2} e^2 \hat{f}^{\mu \nu}(-k)  (k \Theta)^2 \int \ud^4l \ \hat{\Delta}_+(l) P_2(0,-ik \Theta l),
\end{equation*}
and analogously for $h^{\lambda \rho}$. Alternatively, one could modify the action with a nonlocal field strength counterterm.
That the contraction of two fields stemming from the covariant coordinate leads to a linear divergence has already been noticed in~\cite{Gross} in the setting of the modified Feynman rules.

\subsection{$e^1$ from the interaction}
\label{sec:FirstOrder}

If the elementary $n$-point function contributes one power of $e$, then another one must come from the kernels, i.e., we have either $m=2, n=1$ or $m=1, n=2$ in (\ref{eq:2ptCombinatorics}).
We first consider the term (\ref{eq:K1_1}). Using it for the observable involving $f^{\mu \nu}$ and the zeroth order kernel (\ref{eq:K0}) for the second observable, one obtains, using (\ref{eq:3pt}),
\begin{subequations}
\label{subeq:1}
\begin{align}
\label{subeq:A1_1_1_res}
 \Gamma_{\mu \nu \lambda \rho}(k) = 24 (2\pi)^2 g_{\nu \rho} k_\mu k_\lambda & \left\{  \hat{\Delta}_+(k) \int \ud^4 l \ \hat{\Delta}^{(1)}(l) \hat{\Delta}_R(k-l) \sin^2 \tfrac{k \Theta l}{2} \right. \\
\label{subeq:A1_1_2_res}
 & \left. + \hat{\Delta}_A(k) \int \ud^4 l \ \hat{\Delta}_+(l) \hat{\Delta}_+(k-l) \sin^2 \tfrac{k \Theta l}{2}. \right\}
\end{align}
For the term where (\ref{eq:K1_1}) is used in the second observable, one obtains, in the same way, 
\begin{align}
\label{subeq:A1_1_3_res}
 \Gamma_{\mu \nu \lambda \rho}(k) = 24 (2\pi)^2 g_{\nu \rho} k_\mu k_\lambda & \left\{ \hat{\Delta}_+(k) \int \ud^4 l \ \hat{\Delta}^{(1)}(l) \hat{\Delta}_A(k-l) \sin^2 \tfrac{k \Theta l}{2} \right. \\
\label{subeq:A1_1_4_res}
 & \left. + \hat{\Delta}_R(k) \int \ud^4 l \ \hat{\Delta}_+(l) \hat{\Delta}_+(k-l) \sin^2 \tfrac{k \Theta l}{2} \right\}.
\end{align}
\end{subequations}
The terms (\ref{subeq:A1_1_2_res}) and (\ref{subeq:A1_1_3_res}) are finite contributions to the continuous spectrum. 
The loop integrals in~(\ref{subeq:A1_1_1_res}) and~(\ref{subeq:A1_1_4_res}) were already computed in the previous section. Thus, after a field strength renormalization for the planar part, one obtains
\begin{equation}
\label{eq:2ptIR0}
\Gamma_{\mu \nu \lambda \rho}(k) = 12 g_{\nu \rho} k_\mu k_\lambda \hat{\Delta}_+(k) \ln \mu \sqrt{-(k \Theta)^2}
\end{equation}
for the sum of (\ref{subeq:A1_1_1_res}) and (\ref{subeq:A1_1_4_res}).
Here $\mu$ is a mass scale that depends on the renormalization condition.

It remains to discuss the terms involving the term (\ref{eq:K1_2}) of the first order kernel. Using it in the observable with $f^{\mu \nu}$, we obtain
\begin{subequations}
\label{subeq:2}
\begin{align}
 & \Gamma_{\mu \nu \lambda \rho}(k) = - 8 (2\pi)^2 k_\mu k_\lambda \nonumber \\
\label{eq:1stOrderNonlocal}
& \times \left[ \hat{\Delta}_+(k) \int \ud^4 l  \ \hat{\Delta}^{(1)}(l) \hat{\Delta}_R(k-l) \tfrac{\sin^2 \frac{k \Theta l}{2}}{\frac{k \Theta l}{2}} \left\{ 2 (k \Theta)_\rho l_\nu - (k \Theta)_\nu l_\rho + \tfrac{k \Theta l}{2} g_{\nu \rho} \right\} \right. \\
\label{subeq:A1_3_3_res}
& \quad \left. + \hat{\Delta}_A(k) \int \ud^4 l \ \hat{\Delta}_+(l) \hat{\Delta}_+(k-l) \tfrac{\sin^2 \frac{k \Theta l}{2}}{\frac{k \Theta l}{2}} \left\{ 2 (k \Theta)_\rho l_\nu - (k \Theta)_\nu l_\rho + \tfrac{k \Theta l}{2} g_{\nu \rho} \right\} \right].
\end{align}
For the contribution where the term (\ref{eq:K1_2}) is used in the second observable, we obtain
\begin{align}
 & \Gamma_{\mu \nu \lambda \rho}(k) = - 8 (2\pi)^2 k_\mu k_\lambda \nonumber \\
\label{eq:1stOrderNonlocal_A}
 & \times \left[ \hat{\Delta}_+(k) \int \ud^4 l \ \hat{\Delta}^{(1)}(l) \hat{\Delta}_A(k-l) \tfrac{\sin^2 \frac{k \Theta l}{2}}{\frac{k \Theta l}{2}} \left\{ 2 (k \Theta)_\nu l_\rho - (k \Theta)_\rho l_\nu + \tfrac{k \Theta l}{2} g_{\nu \rho} \right\} \right. \\
\label{subeq:A1_3_2_res}
 & \quad \left. + \hat{\Delta}_R(k) \int \ud^4 l \ \hat{\Delta}_+(l) \hat{\Delta}_+(k-l) \tfrac{\sin^2 \frac{k \Theta l}{2}}{\frac{k \Theta l}{2}} \left\{ 2 (k \Theta)_\nu l_\rho - (k \Theta)_\rho l_\nu + \tfrac{k \Theta l}{2} g_{\nu \rho} \right\} \right].
\end{align}
\end{subequations}
The loop integrals in (\ref{subeq:A1_3_3_res}) and (\ref{subeq:A1_3_2_res}) are well-defined and contribute to the continuous spectrum.
The third term in curly brackets in (\ref{eq:1stOrderNonlocal}), respectively (\ref{eq:1stOrderNonlocal_A}), gives rise to a term proportional to~(\ref{subeq:A1_1_1_res}), respectively (\ref{subeq:A1_1_4_res}). For the sum of these, one obtains, cf. (\ref{eq:2ptIR0}),
\begin{equation}
\label{eq:2ptIR0_2}
\Gamma_{\mu \nu \lambda \rho}(k) = - 4 g_{\nu \rho} k_\mu k_\lambda \hat{\Delta}_+(k) \ln \mu \sqrt{-(k \Theta)^2}.
\end{equation}
The first two terms in curly brackets in (\ref{eq:1stOrderNonlocal}) and (\ref{eq:1stOrderNonlocal_A}), are quite unusual, however. Because of the twisting factor in the denominator, it is not possible to split this contribution into a planar and a nonplanar part. Even worse, there is no obvious way to define this integral rigorously.
But we want to compute it at least formally.
The integral over $l$ formally yields an expression of the form $\Sigma_{\nu \rho}(k, \Theta)$. If it is well-defined it should transform properly under Lorentz transformations. 
We consider a timelike $k$ (and later discuss the limit $k^0 \to \betrag{\V{k}}$). Because of Lorentz covariance, it suffices to compute $\Sigma_{\nu \rho}$ for $k = (k_0, \V{0})$ and arbitrary $\Theta $.
For this $k$
\begin{equation}
\label{eq:chiExpression}
 \int \ud^4 l \ \hat{\Delta}^{(1)}(l) \hat{\Delta}_R(k-l) \frac{\sin^2 \frac{k \Theta l}{2}}{\frac{k \Theta l}{2}} l_\nu
\end{equation}
vanishes for $\nu = 0$, because the integrand is antisymmetric under space reflection. The same is true for $\nu = i$ if $e_i$ is perpendicular to $k \Theta $ (now the integrand is antisymmetric under reflection in the direction $k \Theta $). It follows that~(\ref{eq:chiExpression}) is of the form $(k \Theta)_\nu \chi(k)$. We want to compute the function $\chi$ formally.
In a calculation similar to the one performed in Section~\ref{sec:nonplanar}, one obtains, cf. \cite{diss},
\begin{equation}
\label{eq:chi}
 \chi(k) = \frac{2 (2\pi)^{-2}}{(k \Theta)^2} \int_0^\infty \ud l \ \frac{l}{k^2 - 4 l^2 + i \epsilon k_0} \left( 1 - \frac{\sin l \sqrt{-(k \Theta)^2}}{l \sqrt{-(k \Theta)^2}} \right).
\end{equation}
The integral is the formal expression for the difference of the planar and the nonplanar part of the fish graph in the massless $\phi^3$ model, cf. \cite{NCDispRel} and Section~\ref{sec:nonplanar}.
Thus, the sum of the first two terms in curly brackets in~(\ref{eq:1stOrderNonlocal}) is formally given by
\begin{equation}
\label{eq:CovCoorDivergence}
\Gamma_{\mu \nu \lambda \rho}(k) = 8 (2\pi)^2 k_\mu k_\lambda \hat{\Delta}_+(k) \tfrac{(k \Theta)_{\nu} (k \Theta)_{\rho}}{(k \Theta)^2} \left( \Sigma_{\text{pl}}(k) - \Sigma_{\text{np}}(k) \right), 
\end{equation}
where $\Sigma_{\text{pl}}$ and $\Sigma_{\text{np}}$ are the planar and nonplanar part of the self-energy of the massless $\phi^3$ model at second order.
However, $\Sigma_{\text{pl}}(k)$ is nothing but the self-energy known from the commutative case, which is logarithmically divergent. Thus, we have found a divergent quantity that is multiplied with the nonlocal expression $(k \Theta)^{-2}$. Hence, a nonlocal counterterm is unavoidable.

We recall from Section~\ref{sec:nonplanar} that in the difference between the planar and the nonplanar part in (\ref{eq:CovCoorDivergence}) the imaginary part cancels for $k^2 \to 0$. Thus, the sign of the $i \epsilon$-prescription in (\ref{eq:chi}) is not relevant. It follows that for the sum of the first two terms in curly brackets in (\ref{eq:1stOrderNonlocal_A}), one also obtains~(\ref{eq:CovCoorDivergence}).

\subsection{$e^2$ from the interaction}

From the term (\ref{eq:W1}) we obtain
\begin{equation}
\label{eq:2ptIR1}
\Gamma_{\mu \nu \lambda \rho}(k) = - \tfrac{20}{3} g_{\nu \rho} k_\mu k_\lambda \hat{\Delta}_+(k) \ln \mu \sqrt{-(k \Theta)^2}. 
\end{equation}
Finally, the term (\ref{eq:nonplanarIR3}) yields
\begin{equation}
\label{eq:2ptIR2}
\Gamma_{\mu \nu \lambda \rho}(k) = - 4 g_{\nu \rho} \left( 8 \tfrac{\del}{\del m^2} \hat{\Delta}_+(k) - \tfrac{(k \Theta)^2}{3} \hat \Delta_+(k) \right) k_{\mu} k_{\lambda} \tfrac{(k \Theta)_{\nu} (k \Theta)_{\rho}}{(k \Theta)^4}.
\end{equation}

Whether the products of distribution in (\ref{eq:2ptIR1}) and (\ref{eq:2ptIR2}) are well-defined will be discussed in the next section.

\subsection{Summary}

Let us summarize the results of this section. Apart from the contributions to the continuous part of the spectrum, we found the following terms:
\begin{itemize}
\item The term (\ref{eq:divergence1}) came from the contraction of the two photons coming in through the covariant coordinates. It is a nonlocal divergence whose subtraction can be interpreted as a normal ordering of functions of the covariant coordinates.

\item The terms (\ref{eq:2ptIR0}), (\ref{eq:2ptIR0_2}) and (\ref{eq:2ptIR1}) are a momentum-dependent field strength normalization. They sum up to
\begin{equation}
\label{eq:2ptIR1_All}
\Gamma_{\mu \nu \lambda \rho}(k) = -\tfrac{4}{3} g_{\nu \rho} k_\mu k_\lambda \hat{\Delta}_+(k) \ln \mu \sqrt{-(k \Theta)^2}. 
\end{equation}
As we will show in the next section, this expression is well-defined in a certain sense.

\item The term (\ref {eq:CovCoorDivergence}) was obtained formally from (\ref{eq:1stOrderNonlocal}) and (\ref{eq:1stOrderNonlocal_A}). These were the contributions where one power of $e$ came from the covariant coordinate and one from the interacting field. It is a nonlocal expression multiplied with a divergent quantity.

\item The first term in (\ref{eq:2ptIR2}), which arose from the contribution $\Sigma_2$ to the self-energy, is formally a momentum-dependent mass renormalization. As such it was treated in the literature, cf.~\cite{Matusis}. In the next section we will show that this expression is not well-defined even for test functions $\hat{f}$ and $\hat{h}$ vanishing in a neighborhood of the origin. Giving some meaning to this expression introduces nonlocal renormalization ambiguities.
\end{itemize}

\section{Products of distributions and nonlocal renormalization ambiguities}
\label{sec:infrared}

In the preceding section, in (\ref{eq:2ptIR1_All}) and (\ref{eq:2ptIR2}), we encountered products of distributions like $\theta(k_0) \delta(k^2) \ln (- (k \Theta)^2)$ or $\theta(k_0) \delta'(k^2) \frac{1}{(k \Theta)^4}$. Here, we want to discuss the well-definedness of such expressions. We focus on products of the form
\begin{equation}
\label{eq:K}
 \theta(k_0) \delta(k^2) K(-(k \Theta)^2),
\end{equation}
where $K(x)$ might be, e.g., $\ln x$ or $x^{-1}$, i.e., a smooth function apart from a singularity at $x=0$. Products where $\delta(k^2)$ is replaced by $\delta'(k^2)$ can be discussed analogously. Since the tip of the light cone, i.e., the origin $k=0$ might pose additional problems, we (momentarily) exclude it from our considerations by restricting to test functions that vanish in a neighborhood of the origin.

We start by noticing that, as such, the product (\ref{eq:K}) is not well-defined: The wave front set \cite{Hoermander} of $\theta(k_0) \delta(k^2)$, excluding the origin, is
\begin{equation}
\label{eq:WF1}
 \left\{ (k, p_k)| k^2 = 0, k_0 > 0, p_k = \lambda k, \lambda \in \R \setminus \{0\} \right\}.
\end{equation}
Note that the sign of $\lambda$ is not restricted. The wave front set of $K(-(k \Theta)^2)$ is contained in (once more we excluded the origin)
\begin{equation}
\label{eq:WF2}
 \left\{ (k, p_k)| (k \Theta)^2 = 0, k \neq 0, p_k = \lambda k, \lambda \in \R \setminus \{0\} \right\}.
\end{equation}
This might be further restricted by
$\lambda \gtrless 0$, depending on some $i \epsilon$-prescription. For convenience, we restrict our considerations to the case $\Theta = \Theta_0$, cf.~(\ref{eq:sigma_0}). Using Lorentz invariance, the result then applies to all $\Theta \in \Sigma$. Now for $k=(\kappa,0,\pm \kappa,0), \kappa >0$, we have $k^2=0, k_0>0$ and $(k \Theta_0)^2=0$.
Thus, there is an overlap
\begin{equation*}
 N = \{ k \in \R^4 | k_1 = k_3 = 0 \}
\end{equation*}
of the singular supports. Furthermore, the cotangent components of the wave front sets at some fixed $k \in N$ can always add up to zero, even if there is a restriction on the sign of $\lambda$ in the wave front set (\ref{eq:WF2}) of $K(-(k \Theta)^2)$. The reason is that there is no such restriction in the wave front set (\ref{eq:WF1}) of $\theta(k_0) \delta(k^2)$. Hence, the product is not well-defined in the sense of H\"ormander~\cite{Hoermander}.

However, we may take the following point of view. The product~(\ref{eq:K}) is well-defined on test functions $f$ that vanish in a neighborhood of $N$. Explicitly, again for $\Theta = \Theta_0$, we find
\begin{equation*}
 \int \frac{\ud^3 \V k}{2 \betrag{\V{k}}} \ K( 2 \lambda_{\text{nc}}^4 \betrag{ \V{k_\bot} }^2) f(\betrag{\V{k}}, \V{k}),
\end{equation*}
where $\V{k_\bot} = (k_1, 0, k_3)$.
Since $f$ vanishes in a neighborhood of the origin, we may define the test function $\tilde f(\V k) = \frac{1}{2 \betrag{\V k}} f( \betrag{\V k}, \V k)$ and write the above as
\begin{equation*}
 \int \ud^3 \V k \ K( 2 \lambda_{\text{nc}}^4 \betrag{ \V{k_\bot} }^2) \tilde{f}(\V{k}).
\end{equation*}
We may now ask for the possibility to extend the distribution $K(2 \lambda_{\text{nc}}^4 \betrag{ \V{k_\bot} }^2)$ to
\begin{equation*}
\dot N = \{ \V{k} \in \R^3 | \betrag{\V{k_\bot}}=0, \betrag{k_2} > 0 \}.
\end{equation*}
The possibility to do this is governed by its scaling degree
at $\dot N$ \cite{BrunettiFredenhagen}. Since $K$ is only singular at the origin, and the wave front set of $K(2 \lambda_{\text{nc}}^4 \betrag{ \V{k_\bot} }^2)$ is orthogonal to the normal bundle of $\dot N$, the scaling degree can be computed simply by scaling $\V{k_\bot}$. In the case $K(x) = \ln x$, we obtain $0$, while in the case $K(x) = 1/x$, we get $2$. Since the codimension of $\dot N$ in $\R^3$ is $2$, the extension is unique (under the condition that it does not increase the scaling degree) in the first case, but nonunique in the second \cite[Thm.~6.9]{BrunettiFredenhagen}. Thus, in the second case, a nonlocal counterterm is needed. Finally, after extending the distribution to $\dot N$, we can further extend it to the origin. This extension is still unique in the case $K(x) = \ln x$, corresponding to (\ref{eq:2ptIR1_All}) and nonunique in the case $K(x) = 1/x$, which corresponds to the second term in (\ref{eq:2ptIR2}). It is easy to see that the problems become even worse when the product $\theta(k_0) \delta'(k^2) \frac{1}{(k \Theta)^4}$ that occurs in the first term in (\ref{eq:2ptIR2}) is considered.

We have thus shown that~(\ref{eq:2ptIR1_All}) is well-defined, while the extension of the product of distributions in~(\ref{eq:2ptIR2}) has nonlocal ambiguities, i.e., a continuum of renormalization conditions. Note that it does not help to assume only space/space noncommutativity. Then $(k \Theta)^2 = - \lambda_{\text{nc}}^4 \betrag{ \V{k_\bot} }^2$, where $\V{k_\bot}$ is the projection of $k$ on the plane spanned by the noncommuting directions. It follows that $\frac{(k \Theta)_\mu (k \Theta)_\nu}{(k \Theta)^4}$ is proportional to $\frac{1}{\betrag{\V{k_\bot}}^2}$ for $\mu$ and $\nu$ in the noncommuting directions. This is still too singular.

\section{The supersymmetric case}
\label{sec:SUSY}
Now we want to compute the supersymmetric contributions to the full two-point function (\ref{eq:NCQED2pt}) at second order. Our hope is that these cancel the nonlocal divergences found in Section~\ref{sec:Full2pt}. 
We have already seen that the problematic term $\Sigma_2$ in the the self-energy is cancelled by the photino loop.
Here we will show that the contributions coming from the first order kernels (\ref{eq:K1_3},b) cancel the nonlocal divergence (\ref{eq:CovCoorDivergence}). However, the term (\ref{eq:divergence1}), which arose from the contraction of two photons coming in through the covariant coordinate, remains.

\subsection{$e^0$ from the interaction}

As in Section~\ref{sec:FreePart}, the terms where both observables yield one power of $e$, i.e., $m = n = 2$ in \eqref{eq:2ptCombinatorics}, only contribute to the continuous spectrum. Using the two second order kernels (\ref{eq:K2_3},b) in the observable involving $f^{\mu \nu}$ and the elementary four-point function (\ref{eq:W4_SUSY_1}), we find
\begin{multline*}
\Gamma_{\mu \nu \lambda \rho}(k) = \tfrac{i}{2} k_\lambda (k\Theta)_\rho (k \Theta)^\tau \hat \Delta_+(k) \epsilon_{\mu \nu \tau \kappa} \tr (\bar \sigma^\kappa \sigma^\xi) \int \ud^4 l \ l_\xi \hat{\Delta}_+(l) \\
\times \left[ P_2(ik \Theta l, 0) e^{-ik \Theta l} + P_2(0, ik \Theta l) - P_2(-ik \Theta l, 0) e^{ik \Theta l} - P_2(0, -ik \Theta l) \right].
\end{multline*}
As in Section~\ref{sec:FreePart}, the expression in square brackets vanishes. Thus, the divergent contribution (\ref{eq:CovCoorO2_1_result}) is not cancelled. In a sense this had to be expected, since it arose from the contraction of the two photons that came in through the covariant coordinate. Here, we contract the $\lambda_\alpha$ coming from the field strength $W_\alpha$ and the $\bar \lambda_\dotalpha$ from the covariant coordinate.
Thus, the two terms have a different structure and it is not surprising that they do not cancel.

\subsection{$e^1$ from the interaction}

Using the first order kernels (\ref{eq:K1_SUSY}) in the observable involving $f^{\mu \nu}$, the free kernel (\ref{eq:K0}) in the second observable, and the elementary three-point function (\ref{eq:W3_SUSY_1}),
we obtain, using
\begin{equation*}
  \epsilon^{\mu \nu \lambda \xi} g_{\xi \xi'} \epsilon^{\kappa \rho \tau \xi'} = \left( - g^{\mu \kappa} g^{\nu \rho} g^{\lambda \tau} + g^{\mu \kappa} g^{\lambda \rho} g^{\nu \tau} - g^{\lambda \kappa} g^{\mu \rho} g^{\nu \tau}  \right) - \mu \leftrightarrow \nu,
\end{equation*}
cf.\ (\ref{eq:sigmaTrace}), and the antisymmetry of $f^{\mu \nu}$, 
\begin{subequations}
\label{subeq:3}
\begin{align}
\label{subeq:1stSUSY_1}
\Gamma_{\mu \nu \lambda \rho}(k) & = 8 (2\pi)^2 k_\mu k_\lambda \nonumber \\
 & \times \left[ \hat{\Delta}_+(k) \int \ud^4 l \ \hat{\Delta}^{(1)}(l) \hat{\Delta}_R(k-l) \tfrac{\sin^2 \frac{k \Theta l}{2}}{\frac{k \Theta l}{2}} \left\{ (k \Theta)_\rho l_\nu - k \Theta l g_{\nu \rho} \right\} \right. \\
\label{subeq:1stSUSY_2}
 & \quad \left. + \hat{\Delta}_A(k) \int \ud^4 l \ \hat{\Delta}_+(l) \hat{\Delta}_+(k-l) \tfrac{\sin^2 \frac{k \Theta l}{2}}{\frac{k \Theta l}{2}}  \left\{ (k \Theta)_\rho l_\nu - k \Theta l g_{\nu \rho} \right\} \right].
\end{align}
For the terms where (\ref{eq:K1_3},b) are used in the second observable and combined with the corresponding permutation of the elementary three-point function \eqref{eq:W3_SUSY_1}, we find, in the same way,
\begin{align}
\label{subeq:1stSUSY_3}
\Gamma_{\mu \nu \lambda \rho}(k) & = 8 (2\pi)^2 k_\mu k_\lambda \nonumber \\
 & \times \left[ \hat{\Delta}_+(k) \int \ud^4 l \ \hat{\Delta}^{(1)}(l) \hat{\Delta}_A(k-l) \tfrac{\sin^2 \frac{k \Theta l}{2}}{\frac{k \Theta l}{2}} \left\{ (k \Theta)_\nu l_\rho - k \Theta l g_{\nu \rho} \right\} \right. \\
\label{subeq:1stSUSY_4}
 & \quad \left. + \hat{\Delta}_R(k) \int \ud^4 l \ \hat{\Delta}_+(l) \hat{\Delta}_+(k-l) \tfrac{\sin^2 \frac{k \Theta l}{2}}{\frac{k \Theta l}{2}} \left\{ (k \Theta)_\nu l_\rho - k \Theta l g_{\nu \rho} \right\} \right].
\end{align}
\end{subequations}
The terms \eqref{subeq:1stSUSY_2} and \eqref{subeq:1stSUSY_4} are contributions to the continuous spectrum. Adding up \eqref{subeq:1stSUSY_1}, \eqref{subeq:1stSUSY_3}, \eqref{subeq:A1_1_1_res}, \eqref{subeq:A1_1_3_res}, \eqref{eq:1stOrderNonlocal} and \eqref{eq:1stOrderNonlocal_A}, we obtain, using $\Delta = \Delta_R - \Delta_A$,
\begin{equation*}
 \Gamma_{\mu \nu \lambda \rho}(k) = 8 (2\pi)^2 k_\mu k_\lambda \hat{\Delta}_+(k) \int \ud^4 l \ \hat{\Delta}^{(1)}(l) \hat{\Delta}(k-l) \tfrac{\sin^2 \frac{k \Theta l}{2}}{\frac{k \Theta l}{2}} [ (k \Theta)_\nu l_\rho - \nu \leftrightarrow \rho ].
\end{equation*}
The integrand of the loop integral vanishes: $k,l$ and $k-l$ are forced to lie on the light cone, which is only possible if they are parallel, but then $k \Theta l = 0$.
A similar cancellation happens for the contributions to the continuous spectrum. We have thus shown that the contributions to the full two-point function, that involve one kernel of first order, cancel each other. In particular,  the nonlocal divergence (\ref{eq:CovCoorDivergence}) does not appear.

\subsection{$e^2$ from the interaction}
\label{sec:SNCQED_SelfEnergy}

In the supersymmetric case, the self-energy is given by \eqref{eq:Pi_final}. Its contribution to the full two-point function (\ref{eq:NCQED2pt}) is
\begin{equation}
\label{eq:2ptSUSY}
 - 4 e^2 \int \ud^4k \ \hat{f}^{\mu \nu}(-k) \hat{h}^\lambda_{\ \nu}(k) k_\mu k_\lambda \hat{\Delta}_+(k) \ln \mu \sqrt{-(k \Theta)^2},
\end{equation}
which is a momentum-dependent field strength normalization. According to the discussion in Section~\ref{sec:infrared}, it does not require any renormalization. Its effect is studied in the next section.
\begin{remark}
\label{rem:Broken}
In \cite{Carlson} it was shown that in the context of the modified Feynman rules the cancellation of $\Sigma_2$ is not complete if supersymmetry is broken in such a way that the supertrace $M^2$ over the squared masses does not vanish. One then obtains $\Sigma_2(k^2, -s^2) \propto M^2 s^2$. One can convince oneself that the same is true in the Yang-Feldman approach. A term of the above form then gives rise to a product $\theta(k_0) \delta'(k^2) \frac{1}{(k \Theta)^2}$, which, according to the discussion in Section~\ref{sec:infrared}, still has nonlocal ambiguities.
\end{remark}

\section{Acausal effects}
\label{sec:NonlocalEffects}

We recall that the main motivation for the introduction of the noncommutative Minkowski space in \cite{dfr} was the desire to implement space-time uncertainty relations, i.e., some form of nonlocality. It is not surprising that this nonlocality leads to acausal effects, see, e.g., \cite{Causality, WulkenhaarCausality}. However, these are relevant only at the noncommutativity scale and are kinematical in the sense that the nonlocality was put in by hand in the very definition of the noncommutative Minkowski space. Here we want to discuss acausal effects that are created dynamically and are not necessarily limited to the noncommutativity scale.

The distortion of the group velocity discussed in \cite{NCDispRel} was an effect of a momentum-dependent mass normalization. As we saw in the preceding section, in NCSQED, one deals with a momentum-dependent field strength normalization. In the following, we want to show that it leads to a acausal effects.

As can be seen in \eqref{eq:2ptSUSY}, a momentum-dependent field strength normalization multiplies, in momentum space, the free two-point function $\hat{\Delta}_+$. But not only this propagator is modified. Consider a source term $\int \ud^4q \ A^\mu j_\mu$ for the interacting field\footnote{Such a source term is not gauge invariant. Properly, one should couple the field strength to a function of the covariant coordinates. But since the corrections are of order $e$ and our discussion is heuristic, we ignore them.}. We define a new free field ${A^\mu_0}' = A^\mu_0 + \Delta_R \times j^\mu$. The higher order components ${A^\mu_n}'$ of the interacting field are again defined recursively, but now with all lower order components replaced by their primed versions. At first order in $j^\mu$ and zeroth order in $e$, the vacuum expectation value of $\hat{A}_\mu'(k)$ is given by
\[
 (2\pi)^2 \hat{\Delta}_R(k) \hat{\jmath}_\mu(k).
\]
The contribution at second order in $e$ (and again first order in $j^\mu$) is\footnote{Here we assumed that the self-energy is given by \eqref{eq:Pi_final} also for spacelike momenta $k$, cf. Remark~\ref{rem:SigmaSpacelike}.}
\begin{equation*}
 - e^2 (2\pi)^2 \hat \Delta_R(k)  (g_{\mu \nu} k^2 - k_\mu k_\nu ) \left( \ln \mu \sqrt{-(k \Theta)^2} + \order(k^2 (k \Theta)^2) \right) \hat{\Delta}_R(k)  \hat{\jmath}^\nu(k).
\end{equation*}
The part proportional to $k_\mu k_\nu$ is cancelled if one calculates the field strength, so we discard it.
The above then reduces to
\begin{equation*}
 e^2 (2\pi)^2 \hat \Delta_R(k) \left\{ (2\pi)^{-2} \ln \mu \sqrt{-(k \Theta)^2} + \order(k^2 (k \Theta)^2 ) \right\} \hat{\jmath}_\mu(k).
\end{equation*}
We also discard the terms of $\order(k^2 (k \Theta)^2 )$, since they are not propagated (the factor $k^2$ cancels the retarded propagator). We see that up to second order in $e$, the retarded propagator (or equivalently the source $j^\mu$) is modified to
\[
 \hat{\Delta}_R(k) \to \hat{\Delta}_R(k) \left( 1 + e^2 (2\pi)^{-2} \ln \mu \sqrt{-(k \Theta)^2} \right).
\]
The multiplication of the retarded propagator (or the source) in momentum space corresponds to a smearing in position space. In order to examine the strength of this effect, it remains to compute the Fourier transform of $\ln \mu \sqrt{-(k \Theta)^2}$.

We begin by noticing that $\mu$ and the noncommutativity scale are irrelevant. They can be pulled out of the logarithm and can be absorbed in a local field strength renormalization. There is thus no natural scale connected to this effect and one expects a power-law decay. It has been shown in \cite[Eq.~(7.14)]{Guettinger}, that
\begin{equation*}
 (2\pi)^{-2} \int \ud^4k \ \ln \sqrt{k^2} e^{-ikx} = - 2 \pi \delta'(x^2) + c \delta(x).
\end{equation*}
Here $c$ depends on the extension of $\delta'(x^2)$ to the origin. It corresponds to a local wave function renormalization and is thus irrelevant for our us. Thus, the nonlocal kernel is given by
\begin{equation*}
 (2\pi)^{-3} e^2 \delta'((\tilde{\Theta}^{-1} x)^2).
\end{equation*}
Here $\tilde \Theta$ is a descaled version of $\Theta$, such that $\lvert \tilde \Theta \rvert = 1$. Convoluting this with a source $f$, we obtain
\begin{equation*}
 (2\pi)^{-3} e^2 \int \ud^4y \ f(x- \tilde \Theta y) \delta'(y^2).
\end{equation*}
It is easy to see that if $y$ is lightlike, then $\Theta_0 y$ is spacelike or lightlike. Since $\Sigma$ is the orbit of $\Theta_0$ under Lorentz transformations, the same is true for all $\Theta \in \Sigma$. Now assume that $f$ is localized in a region of typical space-time extension $\Delta z$ around the origin. We want to determine its effect at $x$ where $x = \tilde \Theta y$ for $y$ lightlike. We consider $x = (0, \V x)$ (for these $x$ the effect is strongest), i.e., an action at a distance. We then have $\betrag{\V x}^2 = 2 \betrag{ y_0 }^2$. Furthermore, we assume $\betrag{ \V x } \gg \Delta z$.  Thus, using
\begin{equation}
\label{eq:delta'}
 \int \ud^4k \ \hat{f}(k) \theta(k^0) \delta'(k^2) = \int \ud^3 \V k \ \left( \frac{1}{4 \betrag{\V k}^3} \hat{f}(\betrag{\V k}, \V k) - \frac{1}{4 \betrag{\V k}^2} \del_0 \hat{f}(\betrag{\V k}, \V k) \right), 
\end{equation}
we can estimate the relative strength at $x$ to be of the order
\begin{equation*}
 \frac{e^2}{(2\pi)^3} \frac{(\Delta z)^3}{\betrag{\V x}^2} \frac{\betrag{\nabla f(0)}}{\betrag{f(0)}}.
\end{equation*}
Here the second term in (\ref{eq:delta'}) was the leading one.

Unless $f$ is very irregular, the effect seems to be rather weak (recall that we assumed $\betrag{ \V x } \gg \Delta z$), so it might not be in conflict with experimental bounds.
However, we recall that we had to assume unbroken supersymmetry, which is not realistic. As mentioned in Remark~\ref{rem:Broken}, in the case of broken supersymmetry, one will again need nonlocal counterterms.

\section{Conclusion}
\label{sec:Conclusion}

We studied pure NCQED in the Yang-Feldman formalism. We took particular care in using only gauge invariant local quantities (observables). This was achieved by employing covariant coordinates. In this way we computed the two-point function of the interacting field strength at $\order(e^2)$. While in Section~\ref{sec:nonplanar} we recovered the results from calculations in the setting of the modified Feynman rules, we found many additional terms that are due to the covariant coordinates.

We encountered three nonlocal divergences.
The divergent term (\ref{eq:divergence1}) of the two-point function had to be removed by a normal ordering of functions of the covariant coordinates that could also a interpreted as the subtraction of a nonlocal counterterm. While in this case the interpretation is ambiguous, this is not so for the term (\ref{eq:CovCoorDivergence}). It is a term where one power of $e$ comes from the covariant coordinate and one from the interaction. It is a nonlocal expression multiplied by a divergent quantity. Its subtraction can only be interpreted as the subtraction of a nonlocal counterterm. Finally, we provided a new interpretation of the terms (\ref{eq:Hayakawa_Sigma_1}, \ref{eq:Hayakawa_Sigma_2}) already found in the context of the modified Feynman rules. While (\ref{eq:Hayakawa_Sigma_1}), together with the planar contribution, gives rise to a nonlocal contribution to the field strength normalization, the term~(\ref{eq:Hayakawa_Sigma_2}) yields an ill-defined product of distributions that can be extended (renormalized) only upon introducing nonlocal renormalization ambiguities.

In order to study the supersymmetric extension of the model, we defined a supersymmetric version of the covariant coordinates. This lead to a cancellation of the nonlocal divergent term \eqref{eq:CovCoorDivergence}, which came in through the covariant coordinate. Also the term \eqref{eq:Hayakawa_Sigma_2} is cancelled, as already known from the setting of the modified Feynman rules. The remaining nonlocal field strength renormalisation \eqref{eq:2ptSUSY} was then interpreted as giving rise to acausal effects.

Unfortunately, this nice result breaks down if supersymmetry is broken in such a way that the photino acquires a nonvanishing mass. Then it seems necessary to consider nonlocal counterterms if one wants to take NCQED serious as a fundamental theory. This would of course be a major deviation from the standard formalism of quantum field theory. Perhaps this can be justified by considering only counterterms that are, in momentum space, functions of $(k \Theta )^2$,\footnote{Possibly, one could also have a linear dependence on $k^2$ in order to allow for a field strength renormalization. In the present case of NCQED, one would also need counterterms of the forms $\int \ud^4 k \ (\Theta k)^{-4} (k \Theta A(k)) (k \Theta A(-k))$ and $\int \ud^4 k \ k^2 (k \Theta)^{-2} (k \Theta A(k)) (k \Theta A(-k))$, cf. \eqref{eq:W2}.} as proposed in~\cite{LiaoSiboldSpectral}. In the commutative limit, these would be local (albeit divergent). One could then impose the standard dispersion relations as a new renormalization condition. 
At the one-loop level the only remaining effect of the noncommutativity would then be the modification of the two-particle spectrum, which is tiny, cf. \cite{diss}. However, it remains to be investigated whether this can be done consistently at higher loop orders.

\subsection*{Acknowledgments}
The author would like to thank Romeo Brunetti, Claus D\"oscher, Alexander Schenkel and especially Klaus Fredenhagen for valuable comments and discussions.
This work was supported by the German Research Foundation (Deutsche Forschungsgemeinschaft (DFG)) through the Graduiertenkolleg "Zuk\"unftige Entwicklungen in der Teilchenphysik" and the
Institutional Strategy of the University of G\"ottingen. Part of this work was done while visiting the Dipartimento di Matematica of the Universit\`a di Roma ``La Sapienza'' with a grant of the research training network ``Quantum Spaces -- Noncommutative Geometry''. It is a pleasure to thank Sergio Doplicher for his kind hospitality.

\appendix

\section{Spinors and supersymmetry}
\label{app:SUSY}

We mainly use the conventions of~\cite{WessBagger}. However, we use another sign for the metric and for $\sigma^0$. We also changed a sign in the definition of $D_\alpha$ and $\bar D_\dotalpha$.

Weyl spinors are anticommuting. Their indices are raised and lowered with the help of the totally antisymmetric $\epsilon$-tensor:
\begin{gather*}
 \chi^\alpha = \epsilon^{\alpha \beta} \chi_\beta, \quad \chi_\alpha = \epsilon_{\alpha \beta} \chi^\beta, \quad \epsilon^{1 2} = \epsilon_{2 1} = 1, \\
 \bar \chi^\dotalpha = \epsilon^{\dotalpha \dotbeta} \chi_\dotbeta, \quad \bar \chi_\dotalpha = \epsilon_{\dotalpha \dotbeta} \bar \chi^\dotbeta, \quad \epsilon^{\dot 1 \dot 2} = \epsilon_{\dot 2 \dot 1} = 1.
\end{gather*}
Products of Weyl spinors are defined as
\begin{gather*}
 \lambda \chi = \lambda^\alpha \chi_\alpha = - \lambda_\alpha \chi^\alpha = \chi^\alpha \lambda_\alpha = \chi \lambda, \\
 \bar \lambda \bar \chi = \bar \lambda_\dotalpha \bar \chi^\dotalpha = - \bar \lambda^\dotalpha \bar \chi_\dotalpha = \bar \chi_\dotalpha \bar \lambda^\dotalpha = \bar \chi \bar \lambda.
\end{gather*}
The $\sigma$-matrices are defined as
\begin{equation*}
 \sigma^\mu_{\alpha \dotalpha} = ( \1, \sigma^i )_{\alpha \dot \alpha}, \quad \bar \sigma^{\mu \ \dotalpha \alpha} = \epsilon^{\dotalpha \dotbeta} \epsilon^{\alpha \beta} \sigma^\mu_{\beta \dotbeta} = ( \1, - \sigma^i )^{\dotalpha \alpha}.
\end{equation*}  
Here $\sigma^i$ are the usual Pauli matrices. One also defines
\begin{equation*}
 ( \sigma^{\mu \nu} )_\alpha^{\ \beta} = \tfrac{1}{4} \left( \sigma^\mu \bar \sigma^\nu - \sigma^\nu \bar \sigma^\mu \right)_\alpha^{\ \beta}, \quad ( \bar \sigma^{\mu \nu} )^\dotalpha_{\ \dotbeta} = \tfrac{1}{4} \left( \bar \sigma^\mu \sigma^\nu - \bar \sigma^\nu \sigma^\mu \right)^\dotalpha_{\ \dotbeta}.
\end{equation*}
Using the definition above, one finds
\begin{equation*}
 \lambda \sigma^\mu \bar \chi = \lambda^\alpha \sigma^\mu_{\alpha \dotalpha} \bar \chi^\dotalpha = - \epsilon^{\dotalpha \dotbeta} \epsilon^{\alpha \beta} \sigma^\mu_{\alpha \dotalpha} \bar \chi_\dotbeta \lambda_\beta = - \bar \chi \bar \sigma^\mu \lambda.
\end{equation*}
Furthermore, the following identities hold:
\begin{align}
 \sigma^\mu_{\alpha \dotalpha} \bar \sigma^{\nu \ \dotalpha \beta} & = g^{\mu \nu} \delta_\alpha^\beta + 2 ( \sigma^{\mu \nu} )_\alpha^{\ \beta}, \nonumber \\
\label{eq:sigma_trace}
 \tr ( \sigma^\mu \bar \sigma^\nu ) & = 2 g^{\mu \nu}, \\
\label{eq:3sigmas_1}
 \sigma^\mu \bar \sigma^\nu \sigma^\lambda & = - g^{\mu \lambda} \sigma^\nu + g^{\nu \lambda} \sigma^\mu + g^{\mu \nu} \sigma^\lambda + i \epsilon^{\mu \nu \lambda \kappa} \sigma_\kappa, \\
\label{eq:3sigmas_2}
 \bar \sigma^\mu \sigma^\nu \bar \sigma^\lambda & = - g^{\mu \lambda} \bar \sigma^\nu + g^{\nu \lambda} \bar \sigma^\mu + g^{\mu \nu} \bar \sigma^\lambda - i \epsilon^{\mu \nu \lambda \kappa} \bar \sigma_\kappa.
\end{align}
The anticommuting superspace coordinates $\theta$, $\bar \theta$ fulfill
\begin{gather*}
 \theta^\alpha \theta^\beta = - \tfrac{1}{2} \epsilon^{\alpha \beta} \theta^2, \quad \theta_\alpha \theta^\beta = - \tfrac{1}{2} \delta_\alpha^\beta \theta^2, \quad \theta^\alpha \theta_\beta = \tfrac{1}{2} \delta^\alpha_\beta \theta^2, \quad \theta_\alpha \theta_\beta = \tfrac{1}{2} \epsilon_{\alpha \beta} \theta^2, \\
 \bar \theta^\dotalpha \bar \theta^\dotbeta = \tfrac{1}{2} \epsilon^{\dotalpha \dotbeta} \bar \theta^2, \quad \bar \theta_\dotalpha \bar \theta^\dotbeta = \tfrac{1}{2} \delta_\dotalpha^\dotbeta \bar \theta^2, \quad \bar \theta^\dotalpha \bar \theta_\dotbeta = - \tfrac{1}{2} \delta^\dotalpha_\dotbeta \bar \theta^2, \quad \bar \theta_\dotalpha \bar \theta_\dotbeta = - \tfrac{1}{2} \epsilon_{\dotalpha \dotbeta} \bar \theta^2.
\end{gather*}
One defines covariant spinor derivates by
\begin{equation*}
 D_\alpha = \del_\alpha - i \sigma^\mu_{\alpha \dotalpha} \bar \theta^\dotalpha \del_\mu, \quad \bar D_\dotalpha = - \bar \del_\dotalpha + i \theta^\alpha \sigma^\mu_{\alpha \dotalpha} \del_\mu.
\end{equation*}
The partial spinor derivatives are given by
\begin{gather*}
 \del_\alpha \theta^\beta = \delta_\alpha^\beta, \quad   \del_\alpha \theta_\beta = \epsilon_{\beta \alpha}, \quad   \del^\alpha \theta_\beta = \delta_\beta^\alpha, \quad   \del^\alpha \theta^\beta = \epsilon^{\beta \alpha},  \\
 \bar \del_\dotalpha \bar\theta^\dotbeta = \delta_\dotalpha^\dotbeta, \quad   \bar \del_\dotalpha \bar \theta_\dotbeta = \epsilon_{\dotbeta \dotalpha}, \quad   \bar \del^\dotalpha \bar \theta_\dotbeta = \delta_\dotbeta^\dotalpha, \quad  \bar \del^\dotalpha \bar \theta^\dotbeta = \epsilon^{\dotbeta \dotalpha}. 
\end{gather*}
Thus,
\begin{equation}
\label{eq:D_anticomm}
 \{ D_\alpha, \bar D_\dotalpha \} = 2 i \sigma^\mu_{\alpha \dotalpha} \del_\mu.
\end{equation}

\section{The Ward identity}
\label{app:Ward}
We want to explicitly show that in the Yang-Feldman approach, the Ward identity holds at tree level, contrary to the Hamiltonian approach \cite{Ohl}. Adding a term $\bar \psi (i\Dslash - m) \psi$ to the Lagrangean, one obtains the vertices
\begin{center}
\includegraphics{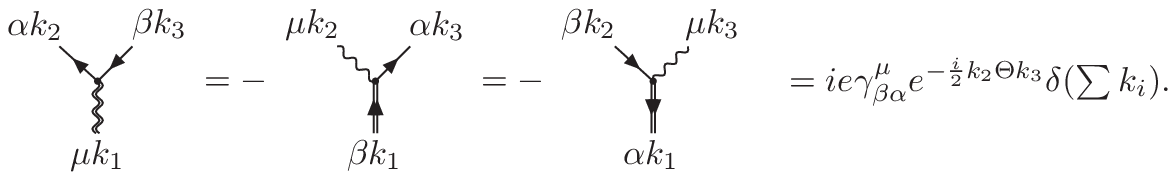}
\end{center}
Here the upward pointing lines correspond to $\psi$, while the downward pointing ones correspond to $\bar \psi$, i.e., to electrons and positrons, for example.
In order to discuss the Ward identity, we consider the vector potential at second order, i.e., the following graphs:
\begin{center}
\includegraphics{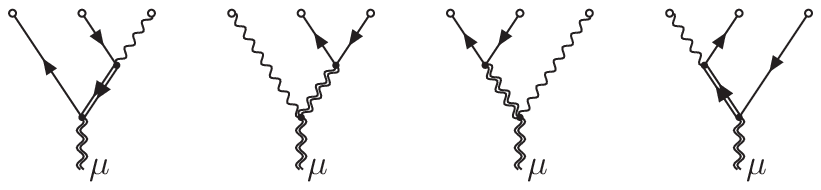}
\end{center}
For the contraction of $\hat{A}_2^\mu(k)$ with $k_\mu$, we obtain
\begin{subequations}
\begin{align}
 k_\mu \hat{A}_2^\mu(k) & = \hat{\Delta}_R(k) \hat{A}_{0 \nu}(k_1) \hat{\psi}_{0 \alpha} (k_2) \hat{\bar \psi}_{0 \beta} (k_3) \delta(k-\sum k_i) \nonumber \\
\label{eq:Ward1}
 & \times \left( - \kslash_{\gamma \alpha} (\kslash - \kslash_2 - m)_{\delta \gamma} \hat{\Delta}_R(k-k_2) \gamma^\nu_{\beta \delta} e^{-\frac{i}{2} k_2 \Theta k} e^{-\frac{i}{2} k_3 \Theta k_1} \right. \\ 
 & \quad + 2i \left( k^\nu (\kslash+\kslash_1)_{\beta \alpha} + \kslash_{\beta \alpha} (k_1-2k)^\nu + \gamma^\nu_{\beta \alpha} (-2k_1+k)\cdot k \right) \nonumber \\
\label{eq:Ward2}
 & \quad \quad \times \sin \tfrac{k_1 \Theta k}{2} e^{-\frac{i}{2} k_2 \Theta k_3} \hat{\Delta}_R(k-k_1) \\ 
\label{eq:Ward3}
 & \quad \left. - \kslash_{\beta \gamma} (-\kslash + \kslash_3 - m)_{\gamma \delta} \hat{\Delta}_R(k-k_3) \gamma^\nu_{\delta \alpha} e^{-\frac{i}{2} k \Theta k_3} e^{-\frac{i}{2} k_1 \Theta k_2} \right)
\end{align}
\end{subequations}
To the first $\kslash$ in the term \eqref{eq:Ward1}, we subtract and add $\kslash_2 - m$. Because of $(\kslash - m)_{\gamma \alpha} \hat{\psi}_{0 \alpha}(k) = 0$, \eqref{eq:Ward1} then reduces to
\[
 (2\pi)^{-2} \gamma^\nu_{\beta \alpha} e^{-\frac{i}{2} k_2 \Theta k} e^{-\frac{i}{2} k_3 \Theta k_1}.
\]
Similarly, using that ${\bar \psi}_0$ is on-shell, the term \eqref{eq:Ward3} reduces to
\[
 - (2\pi)^{-2} \gamma^\nu_{\beta \alpha} e^{-\frac{i}{2} k \Theta k_3} e^{-\frac{i}{2} k_1 \Theta k_2}.
\]
Thus, the sum of the terms \eqref{eq:Ward1} and \eqref{eq:Ward3} yields
\[
 2i (2\pi)^{-2} \gamma^\nu_{\beta \alpha} \sin \tfrac{k_1 \Theta k}{2} e^{-\frac{i}{2} k_2 \Theta k_3}.
\]
Using the on-shell condition for $\psi_0$, ${\bar \psi}_0$ and $A_{0 \nu}$ and the transversality of $A_{0 \nu}$, one finds that the term \eqref{eq:Ward2} reduces to
\[
 - 2i (2\pi)^{-2} \gamma^\nu_{\beta \alpha} \sin \tfrac{k_1 \Theta k}{2} e^{-\frac{i}{2} k_2 \Theta k_3},
\]
so that the the Ward identity at second order tree level is fulfilled. Note that the cubic photon vertex was essential for this cancellation.

\section{Proof of Lemma~\ref{lemma}}
\label{app:lemma}

\begin{proof}
We proceed by induction. The case $N=0$ is obvious. The left hand side can be written as
\begin{align*}
& \sum_{n_0, \dots n_N} \frac{x^{n_0} (x+y_1)^{n_1} \dots (x+y_{N-1})^{n_{N-1}}}{(n_0+ \dots n_N + N)!} \\
  & \qquad \times \sum_{l=0}^{n_N} (x+y_{N-1})^{n_N-l} (y_N-y_{N-1})^l \binom{n_N}{l} \\
= & \sum_{n_0, \dots n_{N-1}} \frac{x^{n_0} (x+y_1)^{n_1} \dots (x+y_{N-2})^{n_{N-2}}}{(n_0+ \dots n_{N-1} + N)!} \\ 
   & \qquad \times  \sum_{l=0}^{n_{N-1}} (x+y_{N-1})^{n_{N-1}-l} (y_N-y_{N-1})^l \sum_{k=l}^{n_{N-1}}  \binom{k}{l} 
\end{align*}
Using $\sum_{k=l}^n \binom{k}{l} = \binom{n+1}{l+1}$ and $\sum_{l=0}^n A^{n-l} B^l \binom{n+1}{l+1} = \frac{1}{B} \left( (A+B)^{n+1} - A^{n+1} \right)$, this can be written as
\begin{align*}
  & \frac{1}{y_N-y_{N-1}} \sum_{n_0, \dots n_{N-1}}  \frac{x^{n_0} (x+y_1)^{n_1} \dots (x+y_{N-2})^{n_{N-2}}}{(n_0+ \dots n_{N-1} + N)!} \\
  & \qquad \times \left( (x+y_N)^{n_{N-1}+1} - (x+y_{N-1})^{n_{N-1}+1} \right) \\
=  & \frac{1}{y_N-y_{N-1}} \sum_{n_0, \dots n_{N-1}} \frac{x^{n_0} (x+y_1)^{n_1} \dots (x+y_{N-2})^{n_{N-2}}}{(n_0+ \dots n_{N-1} + N-1)!} \\
  & \qquad \times \left( (x+y_N)^{n_{N-1}} - (x+y_{N-1})^{n_{N-1}} \right) \\
=  & \frac{e^x}{y_N-y_{N-1}} \sum_{n_1, \dots n_{N-1}} \frac{y_1^{n_1} \dots y_{N-2}^{n_{N-2}}}{\left( \sum_{i=1}^{N-1} (n_i+1) \right)!}  \left( y_N^{n_{N-1}} - y_{N-1}^{n_{N-1}} \right) \\
= & e^x \sum_{n_1, \dots n_N} \frac{1}{\left( \sum_{i=1}^N (n_i+1) \right)!} y_1^{n_1} \dots y_N^{n_N}.
\end{align*}
In the last step we used
\begin{equation*}
 \frac{A^n-B^n}{A-B} = \sum_{k=0}^{n-1} A^k B^{n-1-k}. 
\end{equation*}
\end{proof}


\begin{thebibliography}{99}

\bibitem{dfr}
S.~Doplicher, K.~Fredenhagen and J.~E.~Roberts,
``The Quantum structure of space-time at the Planck scale and quantum fields,''
Commun.\ Math.\ Phys.\  {\bf 172} (1995) 187
[arXiv:hep-th/0303037].

\bibitem{Schomerus}
V.~Schomerus,
``D-branes and deformation quantization,''
JHEP {\bf 9906} (1999) 030
[arXiv:hep-th/9903205].

\bibitem{SW}
N.~Seiberg and E.~Witten,
``String theory and noncommutative geometry,''
JHEP {\bf 9909} (1999) 032
[arXiv:hep-th/9908142].

\bibitem{EField}
N.~Seiberg, L.~Susskind and N.~Toumbas,
``Strings in background electric field, space/time noncommutativity  and a new
noncritical string theory,''
JHEP {\bf 0006} (2000) 021
[arXiv:hep-th/0005040].

\bibitem{LangmannSzabo}
E.~Langmann and R.~J.~Szabo, ``Duality in scalar field theory on noncommutative phase spaces,'' Phys.\ Lett.\ B
{\bf 533} (2002) 168 [arXiv:hep-th/0202039].

H.~Grosse and R.~Wulkenhaar, ``Renormalisation of $\phi^4$-theory on non-commutative $\R^4$ to all orders,''
Lett.\ Math.\ Phys.\  {\bf 71} (2005) 13 [arXiv:hep-th/0403232].

\bibitem{GW2d}
  J.~Zahn,
  ``Divergences in quantum field theory on the noncommutative two-dimensional
  Minkowski space with Grosse-Wulkenhaar potential,''
  arXiv:1005.0541 [hep-th].

\bibitem{p2}
  R.~Gurau, J.~Magnen, V.~Rivasseau and A.~Tanasa,
  ``A translation-invariant renormalizable non-commutative scalar model,''
  Commun.\ Math.\ Phys.\  {\bf 287}, 275 (2009).
  [arXiv:0802.0791 [math-ph]].

\bibitem{twist}
  P.~Aschieri, C.~Blohmann, M.~Dimitrijevic, F.~Meyer, P.~Schupp and J.~Wess,
  ``A gravity theory on noncommutative spaces,''
  Class.\ Quant.\ Grav.\  {\bf 22} (2005) 3511
  [arXiv:hep-th/0504183].

  A.~Schenkel and C.~F.~Uhlemann,
  ``Field Theory on Curved Noncommutative Spacetimes,''
  SIGMA {\bf 6} (2010) 061
  [arXiv:1003.3190 [hep-th]].


\bibitem{Filk}
T.~Filk,
``Divergencies in a field theory on quantum space,''
Phys.\ Lett.\ B {\bf 376} (1996) 53.

\bibitem{chepelev}
  I.~Chepelev and R.~Roiban,
  ``Renormalization of quantum field theories on noncommutative R**d.  I:
  Scalars,''
  JHEP {\bf 0005}, 037 (2000).
  [arXiv:hep-th/9911098].


\bibitem{Minwalla}
S.~Minwalla, M.~Van Raamsdonk and N.~Seiberg,
``Noncommutative perturbative dynamics,''
JHEP {\bf 0002} (2000) 020
[arXiv:hep-th/9912072].


\bibitem{Matusis}
A.~Matusis, L.~Susskind and N.~Toumbas,
``The IR/UV connection in the non-commutative gauge theories,''
JHEP {\bf 0012} (2000) 002
[arXiv:hep-th/0002075].

\bibitem{DoroOWR}
D.~Bahns,
``Locality in Quantum Field Theory on the Noncommutative Minkowski space,''
Report No. 48/2005 in
Oberwolfach Rep. {\bf 2} (2005).


\bibitem{LiaoSibold}
Y.~Liao and K.~Sibold,
``Time-ordered perturbation theory on noncommutative spacetime: Basic  rules,''
Eur.\ Phys.\ J.\ C {\bf 25} (2002) 469
[arXiv:hep-th/0205269].

\bibitem{UVfinite}
D.~Bahns, S.~Doplicher, K.~Fredenhagen and G.~Piacitelli,
``Ultraviolet finite quantum field theory on quantum spacetime,''
Commun.\ Math.\ Phys.\  {\bf 237} (2003) 221
[arXiv:hep-th/0301100]. \\
D.~Bahns,
``Ultraviolet finiteness of the averaged Hamiltonian on the  noncommutative
Minkowski space,''
arXiv:hep-th/0405224.

\bibitem{Ohl}
T.~Ohl, R.~R\"uckl and J.~Zeiner,
``Unitarity of time-like noncommutative gauge theories: The violation of  Ward
identities in time-ordered perturbation theory,''
Nucl.\ Phys.\ B {\bf 676} (2004) 229
[arXiv:hep-th/0309021].

\bibitem{YF}
C.~N.~Yang and D.~Feldman,
``The S Matrix In The Heisenberg Representation,''
Phys.\ Rev.\  {\bf 79} (1950) 972.

\bibitem{BDFP02}
D.~Bahns, S.~Doplicher, K.~Fredenhagen and G.~Piacitelli,
``On the unitarity problem in space/time noncommutative theories,''
Phys.\ Lett.\ B {\bf 533} (2002) 178
[arXiv:hep-th/0201222].

\bibitem{Quasiplanar}
D.~Bahns, S.~Doplicher, K.~Fredenhagen and G.~Piacitelli,
``Field theory on noncommutative spacetimes: Quasiplanar Wick products,''
Phys.\ Rev.\ D {\bf 71} (2005) 025022
[arXiv:hep-th/0408204].

\bibitem{NCDispRel}
C.~D\"oscher and J.~Zahn,
``Dispersion relations in the noncommutative $\phi^3$ and Wess-Zumino model in
the Yang-Feldman formalism,''
Annales Henri Poincar\'e \textbf{10} (2009) 35
[arXiv:hep-th/0605062].

\bibitem{Blaschke}
  D.~N.~Blaschke, E.~Kronberger, R.~I.~P.~Sedmik and M.~Wohlgenannt,
  ``Gauge Theories on Deformed Spaces,''
  arXiv:1004.2127 [hep-th].

\bibitem{Wulkenhaar}
R.~Wulkenhaar,
``Non-renormalizability of $\theta$-expanded noncommutative QED,''
JHEP {\bf 0203} (2002) 024
[arXiv:hep-th/0112248].

\bibitem{Martin}
C.~P.~Mart\'in and D.~S\'anchez-Ruiz,
``The one-loop UV divergent structure of $U(1)$ Yang-Mills theory on
noncommutative $\R^4$,''
Phys.\ Rev.\ Lett.\  {\bf 83} (1999) 476
[arXiv:hep-th/9903077].


\bibitem{hayakawa}
M.~Hayakawa,
``Perturbative analysis on infrared aspects of noncommutative QED on $\R^4$,''
Phys.\ Lett.\ B {\bf 478} (2000) 394
[arXiv:hep-th/9912094]. \\
M.~Hayakawa,
``Perturbative analysis on infrared and ultraviolet aspects of  noncommutative
QED on $\R^4$,''
arXiv:hep-th/9912167.

\bibitem{KhozeTravaglini}
V.~V.~Khoze and G.~Travaglini,
``Wilsonian effective actions and the IR/UV mixing in noncommutative  gauge
theories,''
JHEP {\bf 0101} (2001) 026
[arXiv:hep-th/0011218].

\bibitem{Ruiz}
F.~Ruiz Ruiz,
``Gauge-fixing independence of IR divergences in non-commutative U(1),
perturbative tachyonic instabilities and supersymmetry,''
Phys.\ Lett.\ B {\bf 502} (2001) 274
[arXiv:hep-th/0012171].

\bibitem{UVIRemergent}
  H.~Grosse, H.~Steinacker and M.~Wohlgenannt,
  ``Emergent Gravity, Matrix Models and UV/IR Mixing,''
  JHEP {\bf 0804}, 023 (2008)
  [arXiv:0802.0973 [hep-th]].


\bibitem{GraciaBondia}
V.~Gayral, J.~M.~Gracia-Bond\'ia and F.~Ruiz Ruiz,
``Trouble with space-like noncommutative field theory,''
Phys.\ Lett.\ B {\bf 610} (2005) 141
[arXiv:hep-th/0412235].

\bibitem{Madore}
J.~Madore, S.~Schraml, P.~Schupp and J.~Wess,
``Gauge theory on noncommutative spaces,''
Eur.\ Phys.\ J.\ C {\bf 16} (2000) 161
[arXiv:hep-th/0001203]. \\
D.~Bahns, S.~Doplicher, K.~Fredenhagen and G.~Piacitelli,
``Quantum Geometry on Quantum Spacetime: Distance, Area and Volume Operators,''
arXiv:1005.2130 [hep-th].


\bibitem{NCED}
J.~Zahn,
``Noncommutative electrodynamics with covariant coordinates,''
Phys.\ Rev.\ D {\bf 70} (2004) 107704
[arXiv:hep-th/0405253].

\bibitem{Gross}
D.~J.~Gross, A.~Hashimoto and N.~Itzhaki,
``Observables of non-commutative gauge theories,''
Adv.\ Theor.\ Math.\ Phys.\  {\bf 4} (2000) 893
[arXiv:hep-th/0008075].

\bibitem{Rozali}
M.~Rozali and M.~Van Raamsdonk,
``Gauge invariant correlators in non-commutative gauge theory,''
Nucl.\ Phys.\ B {\bf 608} (2001) 103
[arXiv:hep-th/0012065].

\bibitem{diss}
J.~Zahn,
``Dispersion Relations in Quantum Electrodynamics on the Noncommutative Minkowski Space,''
DESY-THESIS-2006-037 [arXiv:0707.2149 [hep-th]].

 \bibitem{Conservation}
A.~Micu and M.~M.~Sheikh Jabbari,
``Noncommutative $\phi^4$ theory at two loops,''
JHEP {\bf 0101} (2001) 025
[arXiv:hep-th/0008057].

M.~Abou-Zeid and H.~Dorn,
``Comments on the energy-momentum tensor in non-commutative field theories,''
Phys.\ Lett.\ B {\bf 514} (2001) 183
[arXiv:hep-th/0104244].

T.~Pengpan and X.~Xiong,
``A note on the non-commutative Wess-Zumino model,''
Phys.\ Rev.\ D {\bf 63} (2001) 085012
[arXiv:hep-th/0009070].

J.~M.~Grimstrup, B.~Kloibock, L.~Popp, V.~Putz, M.~Schweda and M.~Wickenhauser,
``The energy-momentum tensor in noncommutative gauge field models,''
Int.\ J.\ Mod.\ Phys.\ A {\bf 19} (2004) 5615
[arXiv:hep-th/0210288].

\bibitem{AdLim}
  C.~D\"oscher and J.~Zahn,
  ``Infrared cutoffs and the adiabatic limit in noncommutative spacetime,''
  Phys.\ Rev.\  D {\bf 73}, 045024 (2006).
  [arXiv:hep-th/0512028].

\bibitem{BakLeePark}
D.~Bak, K.~M.~Lee and J.~H.~Park,
``Comments on noncommutative gauge theories,''
Phys.\ Lett.\ B {\bf 501} (2001) 305
[arXiv:hep-th/0011244].

A.~Dhar and S.~R.~Wadia,
``A note on gauge invariant operators in noncommutative gauge theories and the matrix model,''
Phys.\ Lett.\ B {\bf 495} (2000) 413
[arXiv:hep-th/0008144].

\bibitem{Steinacker}
  H.~Steinacker,
  ``Emergent Geometry and Gravity from Matrix Models: an Introduction,''
  Class.\ Quant.\ Grav.\  {\bf 27} (2010) 133001
  [arXiv:1003.4134 [hep-th]].


\bibitem{Itzy}
C.~Itzykson and J.-B.~Zuber,
``Quantum Field Theory,''
McGraw-Hill 1980. 

\bibitem{Gradshteyn}
I.~S.~Gradshteyn and I.~M.~Ryzhik,
``Table of Integrals, Series, and Products,''
Academic Press, 1980.

\bibitem{Hoermander}
L.~H\"ormander,
``The Analysis of Linear Partial Differential Operators,''
Springer, 1990.

\bibitem{BrunettiFredenhagen}
R.~Brunetti and K.~Fredenhagen,
``Microlocal analysis and interacting quantum field theories:  Renormalization
on physical backgrounds,''
Commun.\ Math.\ Phys.\  {\bf 208} (2000) 623
[arXiv:math-ph/9903028].

\bibitem{Carlson}
C.~E.~Carlson, C.~D.~Carone and R.~F.~Lebed,
``Supersymmetric noncommutative QED and Lorentz violation,''
Phys.\ Lett.\ B {\bf 549} (2002) 337
[arXiv:hep-ph/0209077].

\bibitem{Causality}
N.~Seiberg, L.~Susskind and N.~Toumbas,
``Space/time non-commutativity and causality,''
JHEP {\bf 0006} (2000) 044
[arXiv:hep-th/0005015].

\bibitem{WulkenhaarCausality}
H.~Bozkaya, P.~Fischer, H.~Grosse, M.~Pitschmann, V.~Putz, M.~Schweda and R.~Wulkenhaar,
 ``Space/time noncommutative field theories and causality,''
Eur.\ Phys.\ J.\ C {\bf 29} (2003) 133
[arXiv:hep-th/0209253].

\bibitem{Guettinger}
A.~Rieckers and W.~G\"uttinger,
``Spectral Representations of Lorentz Invariant Distributions and Scale Transformation,''
Commun.\ Math.\ Phys.\ {\bf 7} (1968) 190.

\bibitem{LiaoSiboldSpectral}
  Y.~Liao and K.~Sibold,
  ``Spectral representation and dispersion relations in field theory on
  noncommutative space,''
  Phys.\ Lett.\  B {\bf 549} (2002) 352
  [arXiv:hep-th/0209221].

\bibitem{WessBagger}
J.~Wess and J.~Bagger,
``Supersymmetry and supergravity,''
Princeton University Press 1992.


\end{thebibliography}
\end{document}